\documentclass[12pt]{article}

\usepackage{color}
\usepackage{bm}
\usepackage{mathrsfs}
\usepackage{epstopdf}
\usepackage{url}
\usepackage[footnotesize]{caption}
\usepackage{footnote}
\usepackage[caption=false]{subfig}
\usepackage{cite}

\usepackage{textcomp}
\usepackage{dsfont}

\usepackage{amsfonts}
\usepackage{amsmath}
\usepackage{amssymb}
\usepackage{graphicx}
\usepackage{color}
\usepackage[left=1cm,top=1cm,bottom=1cm,right=1cm,nohead,nofoot]{geometry}

\def \be {\begin{equation}}
\def \ee {\end{equation}}
\def \bea {\begin{eqnarray}}
\def \eea {\end{eqnarray}}
\def \nn {\nonumber}

\def \rr {\raise.35ex\hbox{\small $\prime$}\kern-.17em{\mbox{\large $\imath$}}}

\def \dels {\partial\kern-.6em /\kern.1em}
\def \As {{A\kern-.5em / \kern.5em}}
\def \Ds {D\kern-.7em / \kern.5em}

\def \ks {k\kern-.5em /}
\def \ls {l\kern-.5em /}

\reversemarginpar
\newcommand{\hide}[1]{}

\begin{document}
\begin{titlepage}

\begin{center}

\hfill
\vskip .2in

\textbf{\LARGE
Cosmological Implications from $O(D, D)$
\vskip.5cm
}

\vskip .5in
{\large
Chen-Te Ma$^{a,}$\footnote{e-mail address: yefgst@gmail.com} and
Che-Min Shen$^{b,}$\footnote{e-mail address: f97222024@ntu.edu.tw}
\\
\vskip 3mm
}
{\sl
${}^a$
Department of Physics and Center for Theoretical Sciences, \\
${}^b$
Department of Physics and Leung Center for Cosmology and Particle Astrophysics, \\
National Taiwan University, Taipei 10617, Taiwan,
R.O.C.}\\
\vskip 3mm
\vspace{60pt}
\end{center}
\begin{abstract}

Double field theory (DFT) offers a manifest T-duality formulation for massless closed string field theory with both momentum and winding excitations. The gauge symmetry is defined by the generalized Lie derivative which is the extension of the Courant bracket. In this work, we solve and study the properties of FRW (Friedmann-Robertson-Walker) and doubled spherically symmetric metric of DFT.

\end{abstract}

\end{titlepage}

\section{Introduction}
\label{1}
Double field theory (DFT) \cite{Hull:2009mi, Hull:2009zb, Hohm:2010jy, Hohm:2010pp, Siegel:1993xq} is a formulation with doubled dimension. The extra coordinates are motivated by string theory. One question naturally arises, \lq\lq{}Could we know string features from DFT?\rq\rq{}. From common sense, string is one dimensional, but particle is zero dimensional, thus DFT needs more degrees of freedom than usual field theory. One obvious way is to double the coordinates. The extra coordinates are dual to winding modes on massless closed string theory. It turns out that we can have manifest and well-defined T-duality in the case of massless closed string theory. It gives an evidence that we can incorporate string features into field theory from the doubled coordinates this way.

General relativity has at its foundation, semi-Riemannian geometry. From general relativity, we understand that geometry is related to matter. In other words, geometry depends on what kind of objects that we describe because geometry cannot decouple from matter. If the objects are point particles, we can use semi-Riemannian geometry to describe as before. It is a continuous manifold and the distance can be arbitrary small. But this smooth geometry should break down for stringy objects. We possibly need to replace semi-Riemannian geometry by a new geometry. We call this new geometry \lq\lq stringy geometry\rq\rq. Stringy geometry also leads us to some non-geometric problems \cite{Hellerman:2002ax}. We cannot define T-duality in semi-Riemannian geometry so a deeper understanding of stringy geometry \cite{Hohm:2011zr, Andriot:2011uh} should be a good motivation to study double field theory in more details.

One interesting aspect of non-geometric problems is to reformulate the standard ten dimensional supergravity (NS-NS) into new ten dimensional supergravity \cite{Andriot:2011uh, Andriot:2013xca}. This new theory can be obtained from the standard NS-NS Lagrangian by field redefinition from the metric ($g_{mn}$), $b$-field ($b_{mn}$) and dilaton ($\phi$) to the new fields $\tilde{g}_{mn}$, ${\beta}_{mn}$ and $\tilde{\phi}$. This field redefinition leads us to get $Q$- and $R$-flux from the standard three form flux $H$. This new supergravity is called $\beta$-supergravity. Fortunately, this formulation can relate these backgrounds to the four dimensional supergravity. From $\beta$-supergravity, we can study non-geometric flux directly in ten dimension. This is one revolution in double field theory.

Another interesting feature in double field theory is the extension of double field theory to higher derivative $\alpha^{\prime}$ corrections \cite{Hohm:2013jaa}. One related formulation for double field theory is the so-called \lq\lq exceptional field theory\rq\rq \cite{Hohm:2013uia} for eleven dimensional supergravity. Before this was not obvious for $\mbox{E}_{n(n)}$ in eleven dimensional supergravity for each $n$. Especially for $n$=8, we met difficulty of non-closure algebra \cite{Berman:2012vc}. Now, we already knew how to overcome some difficulties to obtain eleven dimensional supergravity by sacrificing some Lonrentz gauge freedoms \cite{Hohm:2013jma}.
For details, we can see some recent reviews in double field theory \cite{Hohm:2013bwa}.

The other related construction is generalized geometry \cite{Gualtieri:2003dx, Hitchin:2004ut}. In generalized geometry, no additional coordinates are introduced. Instead, we consider tangent space and cotangent space together.
A main ingredient in differential geometry is the various notions of derivatives of tensorial elements along the directions of velocity vectors, e.g., the Lie derivatives. The properties of Lie derivatives include Jacobi identity and Leibniz rule are useful in physics. Two important quantities in differential geometry are curvature and torsion. The extension of curvature and torsion in the context of generalized geometry are also discussed \cite{Gualtieri:2007bq}. One of the familiar algebras in generalized geometry is the Courant algebra \cite{Bressler:2005}. Other interesting features of generalized geometry is reduction \cite{Bursztyn:2005vwa}, exceptional generalized geometry (EGG) \cite{Baraglia:2011dg} and supergravity \cite{Hull:2007zu}.

The main point of this paper is to study classical analysis of double field theory since such analysis did not receive much attention; only a few papers worked on this direction \cite{Wu:2013sha, Wu:2013ixa}. The motivation of this work is to understand the structure of solution with respect to manifest $O( D, D)$ structure. The current situation of double field theory is well-defined with strong constraint, but it is not what we expect. Some fluxes do not appear when we do compactification because strong constraint is too restrictive. One direction of relaxing strong constraint is the classical analysis of double field theory-solving equations of motion in double field theory may possibly shed light on relaxing strong constraint. However relaxing constraint is the same as annihilating unwanted solutions. Therefore, solving equations of motion should be a more direct approach to understand this problem. In view of this motivation, we do not use strong constraint to solve equations of motion. In this paper, we focus on FRW and doubled spherically symmetric metric, and perform cosmological analysis. We find that our FRW solutions are unstable against isotropic perturbation. However this only means that the solutions need fine-tuning. In these solutions, we find one solution can potentially  describe inflationary cosmology.

The plan of this paper is to first review
massless closed string of double field theory in Sec. \ref{2}.
Then we show FRW and doubled spherically symmetric solutions of this theory in Sec. \ref{3}.
Finally, we give the summary, conclusions and the possible future directions in Sec. \ref{4}.


\section{Review of DFT for Massless Closed String}
\label{2}

We review the double field theory of massless closed string in this section. At first, We introduce some convenient notations for DFT and then write down the action and gauge symmetry for manifestly background independent \cite{Hohm:2010jy} and generalized metric formulation \cite{Hohm:2010pp}. Finally, we show the derivation of the generalized scalar curvature and Ricci tensor.

\subsection{Basics}
Double field theory is defined on a manifold with doubled dimensions. The fields in DFT depend on the ordinary coordinates $x^i$ and the dual coordinates $\tilde{x}_i$. The dual coordinates are associated with winding excitations. The field components of DFT for massless closed string \cite{Hull:2009mi, Kugo:1992md} are the metric field ($g_{ij}$), antisymmetric field ($b_{ij}$) and scalar dilaton ($d$). This theory is gauge invariant up to cubic order by constraint. (We need to use constraint relation to annihilate all fields and gauge parameters). We have
\bea
\partial_i\tilde{\partial}^i(\mbox{field})=0,
\qquad
\partial_i\tilde{\partial}^i(AB)=0,
\eea
where
\bea
\partial_i=\frac{\partial}{\partial x^i},
\qquad
\tilde{\partial}^i=\frac{\partial}{\partial\tilde{x}_i}.
\eea
The two constraints also imply
\bea
\partial_iA\tilde{\partial}^iB+\tilde{\partial}^iA\partial_iB=0.
\eea
By using the so-called strong constraint above, it can be shown that the fields and gauge parameters in this theory are independent of $\tilde{x}$ \cite{Hull:2009zb}. On the other hand, if we impose the first constraint relation only,
\bea
\partial_i\tilde{\partial}^i(\mbox{field})=0,
\eea
we call it to be weak constraint. But we have
\bea
\partial_i\tilde{\partial}^i\delta(\mbox{field})\neq 0,
\eea
where $\delta$ is the gauge transformation. We need strong constraint to annihilate it. According to the manifest $O(D, D)$ structure in DFT, the $O(D, D)$ invariant weak constraint above can be rewritten as
\bea
\partial^M\partial_M(\mbox{field})=0,
\eea
where the index $\partial_M$ is defined by
\be
{\partial_M }\equiv
 \begin{pmatrix} \,\tilde{\partial}^i \, \\[0.6ex] {\partial_i } \end{pmatrix}
\ee
and $\partial^M=\eta^{MN}\partial_N$. The $\eta$ can also be used to raise and lower the indices for the arbitrary tensor $h$ with respect to $O(D, D)$.
\be
 h= \begin{pmatrix} a& b \\ c& d \end{pmatrix}
\, ,
\, ~~   h^t \eta h = \eta \,,   ~~\eta= \begin{pmatrix} 0& I \\ I& 0 \end{pmatrix}\,,
\ee
 where $a$, $b$, $c$ and $d$ are all $D\times D$ matrices. Then
\be
{\partial^M }\equiv
 \begin{pmatrix} \,\partial_i \, \\[0.6ex] {\tilde{\partial}^i } \end{pmatrix}.
\ee
We can also define $X^M$ to denote coordinates and dual coordinates together by
\be
{X^M}\equiv
 \begin{pmatrix} \,\tilde x_i\, \\[0.6ex] {x^i} \end{pmatrix}.
\ee
Note that $x^i$ and $\tilde{x}_i$ are contravariant, while $x_i$ and $\tilde{x}^i$ are covariant tensor.

\subsection{Action }
In the beginning of this section, we review background independence property of double field theory \cite{Hull:2009mi}. Next, we show the manifestly background independent formulation \cite{Hohm:2010jy}, the generalized metric formulation and the correspondence between manifestly background independent and generalized metric formulation \cite{Hohm:2010pp}.
\subsubsection{Manifestly Background Independence}
The DFT action is
\be
\begin{split}
\label{DFTaction1}
S &=    \hskip-2pt\int [d x d\tilde x] \,
\Bigl[\, \, {1\over 4} e_{ij} D^kD_k e^{ij}  + {1\over 4} (\bar D^j e_{ij})^2
+ {1\over 4} ( D^i e_{ij})^2  - 2 \, d \, D^i \bar D^j e_{ij}
\,-\, 4 \,d \, D^iD_i \, d~~~
\\[0.6ex]
&\hskip15pt + {1\over 4} \,e_{ij}\Bigl( \,(D^i e_{kl})  ( \bar D^j e^{kl} )
 -(D^ie_{kl} )\,( \bar D^l e^{kj})
 - (D^k e^{il} )( \bar D^j e_{kl} )  \Bigr)~
\\[1.0ex]
&\hskip15pt +
{1\over 2} d\,
\Bigl(  (D^ie_{ij})^2  + (\bar D^j e_{ij})^2
+{1\over 2}(D_k e_{ij})^2
 +{1\over 2}( \bar D_k e_{ij} )^2
 + 2e^{ij} (D_i   D^k e_{kj}  + \bar D_j  \bar D^k e_{ik} )
   \Bigr)~  \\[0.9ex]
&\hskip15pt +
4\,e_{ij} d\, D^i \bar D^jd
 + 4 \,d^2 \,D^i D_i\, d~\Bigr]\,.
 \end{split}
\ee
This action up to cubic order is constructed by the scalar dilaton $(d)$ and fluctuation field ($e_{ij}$). (The fluctuation is around the constant background $E_{ij}=G_{ij}+B_{ij}$.) The indices inside this action are raised and lowered by $G_{ij}$. The derivatives $D_i$ and $\bar{D}_i$ are defined by
\bea
D_i\equiv\partial_i-E_{ik}\tilde{\partial}^k,
\qquad
\bar{D}_i\equiv\partial_i+E_{ki}\tilde{\partial}^k.
\eea
 Note that the action we defined here is the same as \cite{Hohm:2010jy}, but not exactly the same as \cite{Hull:2009mi}, where they used $\square\equiv\frac{1}{2}\bigg(D^iD_i+\bar{D}^i\bar{D}_i\bigg)$. We use $\square\equiv\ D^iD_i$ instead. (These choices are the same when constraint is imposed.) From the setup above, we can show background independence without using $\partial^M\partial_M=0$ in the following.
Mathematically, background independence means
\bea
S[E_{ij}, e_{ij}+\chi_{ij}]=S[E_{ij}+\chi_{ij}, e^{\prime}_{ij}=e_{ij}+f_{ij}(\chi, e)],
\eea
where $f_{ij}(\chi, e)$ can redefine $e_{ij}$. That is, the meaning of background independence is to combine $\chi_{ij}$ with the constant background $E_{ij}$. The scalar dilaton does not change here. If $e_{ij}$ vanishes, $e^{\prime}_{ij}$ should also vanish. It implies $f_{ij}$ should contain no $e_{ij}$ independent terms. We can redefine $E_{ij}$ to be $E_{ij}-\chi_{ij}$, and
\bea
e_{ij}=e^{\prime}_{ij}-f_{ij}(\chi, e)
\eea
so we can also have
\bea
S[E_{ij}-\chi_{ij}, e^{\prime}_{ij}+\chi_{ij}-f_{ij}(\chi, e)]=S[E_{ij}, e^{\prime}_{ij}].
\eea
If we expand it to leading order ($e^{\prime}_{ij}\rightarrow e_{ij}$), the restriction of background independence becomes
\bea
S[E_{ij}-\chi_{ij}, e_{ij}+\chi_{ij}-f_{ij}(\chi, e)]=S[E_{ij}, e_{ij}].
\eea
When this action is expanded to cubic order, we can choose
\bea
f_{ij}(\chi, e)=\frac{1}{2}(\chi_i{}^ke_{kj}+\chi^k{}_je_{ik})
\eea
to show this theory is background independence. The background shift ($\delta_B$) is
\bea
\delta_B e_{ij}&=&\chi_{ij}-\frac{1}{2}(\chi_i{}^ke_{kj}+\chi^k{}_je_{ik}),
\nn\\
\delta_B E_{ij}&=&-\chi_{ij},
\nn\\
\delta_B G^{ij}&=&\frac{1}{2}(\chi^{ij}+\chi^{ji}),
\nn\\
\delta_B d&=&0.
\eea
This theory is background independent (under $\delta_B$) up to cubic order. One interesting aspect is that we do not need to use the constraint $\partial^M\partial_M=0$ to prove this for (\ref{DFTaction1}). We can also define a background independent field ${\cal E}_{ij}$.
\bea
{\cal E}_{ij}\equiv g_{ij}+b_{ij}=E_{ij}+e_{ij}+\frac{1}{2}e_{i}{}^ke_{kj}+{\cal O}(e^3),
\qquad
\delta_B E_{ij}={\cal O}(e^2).
\eea
(The constant part is $E_{ij}$.)
Now the background independent variables are ${\cal E}_{ij}$, $d$ and $g^{ij}$. Since the theory has background independence up to cubic order, the action should be rewritten in terms of these background independent variables.
\bea
\label{action_GM}
 \begin{split}\hskip-10pt
  S \ = \ \int \,dx d\tilde{x}~
  e^{-2d}\Big[&
  -\frac{1}{4} \,g^{ik}g^{jl}   \,   {\cal D}^{p}{\cal E}_{kl}\,
  {\cal D}_{p}{\cal E}_{ij}
  +\frac{1}{4}g^{kl} \bigl( {\cal D}^{j}{\cal E}_{ik}
  {\cal D}^{i}{\cal E}_{jl}  + \bar{\cal D}^{j}{\cal E}_{ki}\,
  \bar{\cal D}^{i}{\cal E}_{lj} \bigr)~
\\ &    + \bigl( {\cal D}^{i}\hskip-1.5pt d~\bar{\cal D}^{j}{\cal E}_{ij}
 +\bar{{\cal D}}^{i}\hskip-1.5pt d~{\cal D}^{j}{\cal E}_{ji}\bigr)
 +4{\cal D}^{i}\hskip-1.5pt d \,{\cal D}_{i}d ~\Big]\;,
 \end{split}
 \eea
where
\bea
{\cal D}_i\equiv\partial_i-{\cal E}_{ik}\tilde{\partial}^k,
\qquad
\bar{\cal D}_i\equiv\partial_i+{\cal E}_{ki}\tilde{\partial}^k.
\eea
Then we say this action is manifestly background independent because all elements now are background independent variables. Note that now the indices are raised and lowered by $g^{ij}$.
The constraint $\partial_M A\partial^M B=0$ can be rewritten as
\bea
\label{sconstraint}
{\cal D}^i A{\cal D}_i B=\bar{\cal D}^i A\bar{\cal D}_i B.
\eea
Furthermore, $\partial_M\partial^MA=0$ can also be rewritten as
\be
{\cal E}_{ij} \left(\bar{\cal D}^{j}\bar{\cal D}^{i}
-{\cal D}^{i}{\cal D}^{j}
-\bar{\cal D}^{j}{\cal D}^{i}+
{\cal D}^{i}\bar{\cal D}^{j}
\right) A=0\;
\ee
by using the relations
 \bea
  \partial_{i}=\frac{1}{2}\left({\cal E}_{ji}{\cal D}^{j}+{\cal E}_{ij}\bar{\cal D}^{j}
  \right)\;, \qquad
  \tilde{\partial}^{i}=\frac{1}{2}\left(-{\cal D}^{i}+\bar{\cal D}^{i}\right)\;.
 \eea
We make an additional remark that this theory also has $\mathbb{Z}_2$ symmetry. This means exchanging the
indices in ${\cal E}$, the barred derivatives with unbarred ones, leave the action invariant \cite{Hull:2009mi}. It can be checked by using (\ref{sconstraint}) to show $\mathbb{Z}_2$ symmetry for the first and last term of (\ref{action_GM}).

\subsubsection{Generalized Metric Formulation}
In this section, we first introduce the generalized metric ${\cal H}_{MN}$.
   \be
  {\cal H}~ \equiv ~ {\cal H}^{\bullet\,\bullet}  \,,
  \ee
\be
  {\cal H} \ = \
  \begin{pmatrix}    g-bg^{-1}b & bg^{-1}\\[0.5ex]
  -g^{-1}b & g^{-1}\end{pmatrix}\, .
 \ee
This matrix is a symmetric matrix with $O(D, D)$ symmetry,
 \be
 {\cal H}\,\eta \,{\cal H}=\eta\;.
  \ee
The inverse of it is
\be
{\cal H}^{-1} =  \eta {\cal H} \eta\,.
\ee

\be
  {\cal H}^{-1}~ \equiv ~ {\cal H}_{\bullet\,\bullet}  \,\ = \ \left({\cal H}^{MN}\right)^{-1} \ = \
  \begin{pmatrix}    g^{-1} & -g^{-1}b\\[0.5ex]
  bg^{-1} & g-bg^{-1}b\end{pmatrix}\;.
  \ee
${\cal H}$ and ${\cal H}^{-1}$ are both symmetric matrix and inverse to each other as usual metric.
In the following, we show how to rewrite the action in terms of generalized metric and scalar dilaton. Because this theory has manifest $O(D, D)$ structure and $\mathbb{Z}_2$  symmetry as we discussed in the previous section, the construction now should also respect these symmetry. Consider the $\mathbb{Z}_2$ symmetry,
\bea
b_{ij}\rightarrow -b_{ij},
\qquad
\tilde{\partial}\rightarrow -\tilde{\partial}.
\eea
This implies
\bea
{\cal E}_{ij}\rightarrow{\cal E}_{ji},
\eea
and $\tilde{\partial}\rightarrow -\tilde{\partial}$ can be rewritten as
\be
\partial_M\rightarrow  Z\, \partial_M \,, ~~~\hbox{with} ~~
Z= \begin{pmatrix} -1 & 0 \\ \phantom{-}0& 1 \end{pmatrix}\,,
\ee
where $Z$ satisfies
\be
Z = Z^t \,,\qquad  Z^2 = 1\,.
\ee
Under the transformation $b_{ij} \rightarrow - b_{ij}$, the off-diagonal matrices in ${\cal H}^{MN}$
change sign. That means
\be
{\cal H}^{MN}  \to  Z {\cal H}^{MN} Z\,, \qquad
{\cal H}_{MN}  \to  Z {\cal H}_{MN} Z \,.
\ee
One should carefully note that $Z$ is not an $O(D,D)$ matrix, i.e.,
\be
\eta^{MN}  \not=  Z \,\eta ^{MN} Z\,, \qquad
\eta_{MN}  \not=  Z \,\eta_{MN} Z \,.
\ee
By using $\partial_M,  {\cal H}^{MN}$, $ {\cal H}_{MN}$ and $d$, we can construct the action with respect to the gauge symmetry (restricted to the strong constraint) by all possible terms within third order up to a boundary term. The action is
\bea
\label{acg}
S &=& \int dx \ d\tilde x  \
   e^{-2d}\Big(\frac{1}{8}{\cal H}^{MN}\partial_{M}{\cal H}^{KL}
  \partial_{N}{\cal H}_{KL}-\frac{1}{2}
  \,{\cal H}^{MN}\partial_{N}{\cal H}^{KL}\partial_{L}
  {\cal H}_{MK}
\nn\\
  &&-2\partial_{M}d\partial_{N}{\cal H}^{MN}+4{\cal H}^{MN}\,\partial_{M}d
  \partial_{N}d \Big).
 \eea
By using the conditions below,
\bea
  \int [dxd\tilde x]\ e^{-2d}\bigg(4{\cal H}^{MN}\partial_{M}d\partial_{N}d\bigg)
  &=&\int [dxd\tilde x]\ e^{-2d}\bigg(4g^{ij}\bar{\cal D}_{i}d\,\bar{\cal D}_{j}d\bigg),
\nn\\
  \int [dxd\tilde x]\ e^{-2d}\bigg(-2\,\partial_{M}d\,\partial_{N}{\cal H}^{MN}\bigg) \ &=&
  \int [dxd\tilde x]\ e^{-2d}\bigg(-2 \,\tilde\partial^k d \, \bar {\cal D}_j {\cal E}_{ki} \, g^{ij}
  -2  \, \bar {\cal D}_i d \, \tilde\partial^k {\cal E}_{kj} \, g^{ij}
  \nn\\
  &&- 2\,
  \bar{\cal D}_i d \, \bar{\cal D}_j g^{ij}\bigg),
\nn\\
 \int [dxd\tilde x]\ e^{-2d}\bigg(-2\,\partial_{M}d\,\partial_{N}{\cal H}^{MN}\bigg) &=&\int [dxd\tilde x]\
e^{-2d}\bigg(  g^{ij}g^{kl}\left({\cal D}_{l}d\,\bar{\cal D}_{j}{\cal E}_{ki}+\bar{\cal D}_{i}d\,
  {\cal D}_{l}{\cal E}_{kj}\right)\bigg),
\nn\\
\int [dxd\tilde x]\ e^{-2d}\bigg(\frac{1}{8}{\cal H}^{MN}\partial_{M}{\cal H}^{KL}
  \,\partial_{N}{\cal H}_{KL}\bigg)&=&\int [dxd\tilde x]\
e^{-2d}\bigg(  -\frac{1}{4} \,g^{ik}g^{jl}   \,   {\cal D}^{p}{\cal E}_{kl}\,
  {\cal D}_{p}{\cal E}_{ij}\bigg)\,,
\nn\\
\int [dxd\tilde x]\ e^{-2d}\bigg(-\frac{1}{2}{\cal H}^{MN}\partial_{N}{\cal H}^{KL}\,\partial_{L}
  {\cal H}_{MK}\bigg) &=&\int [dxd\tilde x]\ e^{-2d}\bigg(\frac{1}{4}g^{kl} \bigl( {\cal D}^{j}{\cal E}_{ik}
  {\cal D}^{i}{\cal E}_{jl}  + \bar{\cal D}^{j}{\cal E}_{ki}\,
  \bar{\cal D}^{i}{\cal E}_{lj} \bigr)\,\bigg),
  \nn
\eea
we can show the action (\ref{acg}) is indeed the same as the original action (\ref{action_GM}).

If we assume all fields in the action are independent of $\tilde{x}$, it can be checked easily that the action would reduce to the familiar action
\bea
\int dx\ \sqrt{-g}e^{-2\phi}\bigg( R+4(\partial\phi)^2-\frac{1}{12}H^2\bigg),
\eea
where $\phi$ is the dilaton satisfying $\sqrt{-g}e^{-2\phi}=e^{-2d}$, $R$ is the Ricci scalar and $H=db$ is the the three form field strength. Here $b$ is  the two form potential corresponding to the field strength $H$.

\subsection{Gauge Symmetry, Generalized Lie Derivative and Brackets}
In this section, we review the gauge symmetry for the two different formulations. Furthermore, we also introduce the generalized Lie derivative, $C$-, and $D$-brackets. Finally, we show how to derive Courant and Dorfman bracket from $C$- and $D$-brackets by using constraint.

The gauge transformation in manifestly background independent formulation is
\bea
\label{gauge tran_BI}
\delta_{\xi}{\cal E}_{ij}&=&{\cal D}_i\tilde{\xi}_j-\bar{{\cal D}}_j\tilde{\xi}_i+\xi^M\partial_M{\cal E}_{ij}+{\cal D}_i\xi^k{\cal E}_{kj}+\bar{{\cal D}}_{j}\xi^k{\cal E}_{ik},
\nn\\
\delta_{\xi} d&=&-\frac{1}{2}\partial_M\xi^M+\xi^M\partial_M d.
\eea
The above gauge transformation is based on strong constraint.
The gauge transformation has closure property.
\bea
[\delta_{\xi_1}, \delta_{\xi_2}]=-\delta_{[\xi_1, \xi_2]_C},
\eea
where the $C$-bracket is defined by
\bea
[\xi_1, \xi_2]_C^M=\xi_1^N\partial_N\xi_2^M-\xi_2^N\partial_N\xi_1^M-\frac{1}{2}\eta^{MN}\eta_{PQ}\xi_1^P\partial_N\xi_2^Q+\frac{1}{2}\eta^{MN}\eta_{PQ}\xi_2^P\partial_N\xi_1^Q.
\eea
We get
\bea
\delta_{\xi}^{(0)}S^{(0)}&=&0,
\nn\\
\delta_{\xi}^{(1)}S^{(2)}&=&0,
\nn\\
\delta_{\xi}^{(0)}S^{(1)}+\delta_{\xi}^{(1)}S^{(0)}&=&0,
\nn\\
\delta_{\xi}^{(1)}S^{(1)}+\delta_{\xi}^{(0)}S^{(2)}&=&0.
\eea
(The superscript means the number of $\tilde{\partial}$.) $\delta^{(0)}$, $\delta^{(1)}$ and $S^{(0)}$, $S^{(2)}$ are T-dual to each other. The second and fourth relations are related to the first and third relations by T-duality. We only need to check the first and third relations.

In the generalized metric formulation, the gauge transformation of the scalar dilaton is the same as the second equation of (\ref{gauge tran_BI}). The gauge transformation of the generalized metric is
\bea
\delta_{\xi} {\cal H}^{MN}=\xi^P\partial_P{\cal H}^{MN}+(\partial^M\xi_p-\partial_P\xi^M){\cal H}^{PN}+(\partial^N\xi_P-\partial_P\xi^N){\cal H}^{MP}.
\eea
Then we can define the generalized Lie derivative
\bea
\hat{\cal L}_{\xi}{\cal H}^{MN}\equiv\delta_{\xi}{\cal H}^{MN},
\eea
which satisfies Leibniz rule and
\bea
\hat{\cal L}_{\xi+\eta^{-1}\partial\chi}A=\hat{\cal L}_{\xi}A,
\qquad
\hat{\cal L}_{\xi}\eta_{MN}=0,
\qquad
\hat{\cal L}_{\xi}\eta^{MN}=0,
\qquad
\hat{\cal L}_{\xi}\delta_M{}^N=0.
\eea
Ordinary Lie derivative acting on $\eta$ is not zero, but now the generalized Lie derivative acting on $\eta$ \emph{is} zero.
The gauge transformation is also compatible with manifest $O(D, D)$ structure.
\bea
&&\hat{\cal L}_{\xi}({\cal H}\eta{\cal H})=0=(\hat{\cal L}_{\xi}{\cal H})\eta{\cal H}+{\cal H}\eta(\hat{\cal L}_{\xi}{\cal H}),
\nn\\
&&\delta_{\xi}({\cal H}\eta{\cal H})=0=(\delta_{\xi}{\cal H})\eta{\cal H}+{\cal H}(\delta_{\xi}{\cal H}).
\eea
The gauge algebra is closed by assuming strong constraint.
\bea
[\hat{\cal L}_{\xi_1}, \hat{\cal L}_{\xi_2}]=\hat{\cal L}_{[\xi_1, \xi_2]_C}.
\eea
This also implies
\bea
[\delta_{\xi_1}, \delta_{\xi_2}]=-\delta_{[\xi_1, \xi_2]_C}.
\eea
Be careful that the signs are different in the above two relations. This is due to the fact that generalized Lie derivative is an operator, but the gauge transformation is a variation, so in general, $\delta_{\xi}\neq\hat{\cal L}$. The $D$-bracket is defined by
\bea
[A, B]_D\equiv\hat{\cal L}_A B.
\eea
The relation between $D$-bracket and $C$-bracket is
\bea
[A, B]_D^M=[A, B]_C^M+\frac{1}{2}\partial^M(B^N A_N),
\eea
where $A$ and $B$ are generalized vectors.
Finally, we show how to obtain Courant bracket from $C$-bracket if we assume all parameters are independent of $\tilde{x}$ \cite{Hull:2009zb}.
\bea
\lbrack\xi_1, \xi_2\rbrack_C^i&=&\xi_1^j\partial_j\xi_2^i-\xi_2^j\partial_j\xi_1^i=({\cal L}_{\xi_1}\xi_2)^i=([\xi_1, \xi_2])^i,
\nn\\
\lbrack\xi_1, \xi_2\rbrack_{Ci}&=&\xi^j_1\partial_j\tilde{\xi}_{2i}-\xi^j_2\partial_j\tilde{\xi}_{1i}
-\frac{1}{2}(\xi_1^j\partial_i\tilde{\xi}_{2j}-\tilde{\xi}_{2j}\partial_i\xi^j_1)
+\frac{1}{2}(\xi_2^j\partial_i\tilde{\xi}_{1j}-\tilde{\xi}_{1j}\partial_i\xi^j_2)
\nn\\
&=&
\xi^j_1\partial_j\tilde{\xi}_{2i}-\xi^j_2\partial_j\tilde{\xi}_{1i}
+(\partial_i\xi_1^j)\tilde{\xi}_{2j}-\frac{1}{2}\partial_i(\xi_1^j\tilde{\xi}_{2j})
-(\partial_i\xi_2^j)\tilde{\xi}_{1j}+\frac{1}{2}\partial_i(\xi_2^j\tilde{\xi}_{1j})
\nn\\
&=&\bigg({\cal L}_{\xi_1}\tilde{\xi}_2-\frac{1}{2}d(i_{\xi_1}\tilde{\xi}_2)\bigg)_i
-\bigg({\cal L}_{\xi_2}\tilde{\xi}_1-\frac{1}{2}d(i_{\xi_2}\tilde{\xi}_1)\bigg)_i.
\eea
It is the same as $[\xi_1+\tilde{\xi}_1, \xi_2+\tilde{\xi}_2]_{\mbox{Cour}}$.
\bea
[A+\alpha, B+\beta]_{\mbox{Cour}}=[A, B]+{\cal L}_A\beta-{\cal L}_B\alpha-\frac{1}{2}d(i_A\beta-i_B\alpha),
\eea
where $A$, $B$ are vector fields, and $\alpha$, $\beta$ are one-form fields.
Similarly, we can obtain Dorfman bracket \cite{Gualtieri:2003dx} from $D$-bracket.
\bea
[A+\alpha, B+\beta]_{\mbox{Dor}}=[A, B]+{\cal L}_A\beta-i_Bd\alpha.
\eea
For consistent notation, we use the above form to express Dorfman bracket instead of the conventional way $(A+\alpha)\circ(B+\beta)$. $D$-bracket also has Jacobi identity as Dorfman bracket.
\bea
[A, [B, C]_D]_D=[[A, B]_D, C]_D+ [B, [A, C]_D]_D.
\eea
But it is not antisymmetric. In additional, $C$-bracket does not satisfy Jacobi identity, but is antisymmetric. Therefore $C$ and $D$-bracket are \emph{not} Lie algebra.
\subsection{Generalized Scalar Curvature and Ricci Tensor}
In this section, we show how to define generalized scalar curvature and Ricci tensor in generalized metric formulation. The generalized scalar curvature is defined by the equations of motion of scalar dilation.
\bea
{\cal R}&\equiv& 4{\cal H}^{MN}\partial_M\partial_N d-\partial_M\partial_N{\cal H}^{MN}-4{\cal H}^{MN}\partial_M d\partial_N d+4\partial_M{\cal H}^{MN}\partial_N d
\nn\\
&&+\frac{1}{8}{\cal H}^{MN}\partial_M{\cal H}^{KL}\partial_N{\cal H}_{KL}
-\frac{1}{2}{\cal H}^{MN}\partial_M{\cal H}^{KL}\partial_K{\cal H}_{NL}.
\eea
It satisfies
\bea
\delta_{\xi}{\cal R}=\xi^M\partial_M{\cal R}.
\eea
By variating the $O(D, D)$ tensor $\cal{H}^{MN}$, we can obtain the other equation of motions used to define the generalized Ricci tensor. We need to add one constraint to the action,
\bea
\lambda_{MN}({\cal H}\eta{\cal H}-\eta)^{MN}.
\eea
By the variation of ${\cal H}^{MN}$, we can get
\bea
{\cal K}_{MN}+(\lambda S+S^{t}\lambda)_{MN}=0,
\eea
where
\bea
&&\lambda\equiv\lambda_{\bullet\bullet},
\qquad S\equiv{\cal H}\eta,
\qquad
S^2=1,
\nn\\
&&{\cal K}_{MN}\equiv\frac{1}{8}\partial_M{\cal H}^{KL}\partial_N{\cal H}_{KL}-\frac{1}{4}(\partial_L-2(\partial_L d))({\cal H}^{LK}\partial_K{\cal H}_{MN})+2\partial_M\partial_N d
\nn\\
&&\ \ \ \ \ \ \ \ \ \  -\frac{1}{2}\partial_{(M}{\cal H}^{KL}\partial_L{\cal H}_{N)K}+\frac{1}{2}(\partial_L-2(\partial_L d))({\cal H}^{KL}\partial_{(M}{\cal H}_{N)K}+{\cal H}^K{}_{(M}\partial_K{\cal H}^L{}_{N)}).
\nn
\eea
We also have
\bea
(S^{t}KS)_{MN}+(S^{t}\lambda+\lambda S)_{MN}=0.
\eea
It implies
\bea
{\cal K}_{MN}-(S^{t}KS)_{MN}=0.
\eea
Then the generalized Ricci tensor is defined by
\bea
{\cal R}_{MN}\equiv \frac{1}{2}{\cal K}_{MN}-\frac{1}{2}(S^{t}KS)_{MN}.
\eea


\section{Solutions for the DFT}
\label{3}
In this section, we show FRW solutions and discuss the stability issue. Then we also show doubled spherically symmetric metric solutions.
We first start with the action
\bea
S &=& \int dx \ d\tilde x  \
   e^{-2d}\Big(\frac{1}{8}{\cal H}^{MN}\partial_{M}{\cal H}^{KL}
  \partial_{N}{\cal H}_{KL}-\frac{1}{2}
  \,{\cal H}^{MN}\partial_{N}{\cal H}^{KL}\partial_{L}
  {\cal H}_{MK}
\nn\\
  &&-2\partial_{M}d\partial_{N}{\cal H}^{MN}+4{\cal H}^{MN}\,\partial_{M}d
  \partial_{N}d + V(d) \Big),
 \eea
where the potential $V(d)=V_0$. In this case, we do not lose diffeomorphism. If potential depends on scalar dilaton, however  probably breaks diffeomorphism. That is why we do not include this case in our discussion.

The equation of motions, i.e. the generalized scalar curvature and Ricci tensor are
\bea
{\cal R}&\equiv& 4{\cal H}^{MN}\partial_M\partial_N d-\partial_M\partial_N{\cal H}^{MN}-4{\cal H}^{MN}\partial_M d\partial_N d+4\partial_M{\cal H}^{MN}\partial_N d
\nn\\
&&+\frac{1}{8}{\cal H}^{MN}\partial_M{\cal H}^{KL}\partial_N{\cal H}_{KL}
-\frac{1}{2}{\cal H}^{MN}\partial_M{\cal H}^{KL}\partial_K{\cal H}_{NL}+V_0 = 0,
\nn\\
{\cal R}_{MN}&\equiv& \frac{1}{2}{\cal K}_{MN}-\frac{1}{2}(S^{t}KS)_{MN} = 0.
\eea

\subsection{FRW metric}
In order to solve FRW metric solution, we first assume the D-dim metric field $g_{ij} = \mbox{diag} \bigg( -1, a(t,\tilde{t}), \cdots, a(t,\tilde{t}) \bigg)$ and $b_{ij} = 0$, and then the EOM are

\bea
{\cal R}&=&0 \Rightarrow (4d^{\prime\prime} - 4d^{\prime 2} -(D-1)\tilde{H}^2) + (4\ddot{d} - 4\dot{d}^2 -(D-1) H^2) + V_0 = 0,
\nn\\
{\cal R}_{tt}&=&0 \Rightarrow (2d^{\prime\prime} -(D-1)\tilde{H}^2) - (2\ddot{d} -(D-1) H^2) = 0,
\nn\\
{\cal R}_{ii}&=&0 \Rightarrow ( \tilde{H}^{\prime} -2\tilde{H}d^{\prime} ) + ( \dot{H} -2H\dot{d} )  = 0,
\eea
where $H \equiv \frac{\dot{a}}{a}$, $\tilde{H} \equiv \frac{a^{\prime}}{a}$. We define  ${}^{\prime} \equiv \partial_{\tilde{t}}$ and $ \dot{} \equiv \partial_{t}$ when considering FRW metric.
In the following, we consider three solutions to FRW metric. The first solution is found in \cite{Wu:2013sha} and we also found another two nontrivial FRW metric solutions.

\subsubsection{First FRW Metric Solution}
In the first case, we assume $V_0 = 0$ and
\bea
\label{1stFRWt}
4\ddot{d} - 4\dot{d}^2 -(D-1) H^2 &=& 0,
\nn\\
2\ddot{d} -(D-1) H^2 &=& 0,
\nn\\
\dot{H} -2H\dot{d}  &=& 0.
\eea
That means
\bea
4d^{\prime\prime} - 4d^{\prime 2} -(D-1)\tilde{H}^2 &=& 0,
\nn\\
2d^{\prime\prime} -(D-1)\tilde{H}^2 &=& 0,
\nn\\
\tilde{H}^{\prime} -2\tilde{H}d^{\prime} &=& 0,
\eea
and the solutions of the EOM above are
\bea
a(t, \tilde{t} ) = a_0 A(t)\cdot\tilde{A}(\tilde{t}) ,\qquad
d(t, \tilde{t} ) = -\frac{1}{2} \ln\bigg((t+t_0)(\tilde{t}+\tilde{t}_0) \bigg) + d_0,
\eea
where

\bea
A(t)=(t+t_0)^{ \frac{1}{\sqrt{D-1}}} \qquad \mbox{or} \qquad (t+t_0)^{ - \frac{1}{\sqrt{D-1}}},
\nn\\
\tilde{A}(\tilde{t})=(\tilde{t}+\tilde{t}_0)^{ \frac{1}{\sqrt{D-1}}} \qquad \mbox{or} \qquad (\tilde{t}+\tilde{t}_0)^{ - \frac{1}{\sqrt{D-1}}},
\eea
\bea
(t+t_0)(\tilde{t}+\tilde{t}_0)\neq0,
\eea
and $t_0$, $\tilde{t}_0$, $a_0$, $d_0$ are constants.
Now the scale factor $a(t, \tilde{t} ) = a_0 A(t)\cdot\tilde{A}(\tilde{t})$ depends on the two time coordinates $t$, $\tilde{t}$, and $\tilde{A}(\tilde{t})$ is not periodic. This solution corresponds to a world with two time coordinates and the physical meaning of it is somehow nontrivial.

However, as the work in \cite{Wu:2013sha}, in this case, we can compactify the dual time coordinate, and then obtain the EOM (\ref{1stFRWt}) which depend on $t$ only.  The solution is as follows:
\bea
a_\pm(t) = a_0\cdot (t+t_0)^{\pm \frac{1}{ \sqrt{D-1}}},\qquad
d(t) = -\frac{1}{2}\bigg( \ln(t+t_0) \bigg) + d_0,
\eea
where $a_0$, $d_0$, $t_0$ are constants, and $t+t_0\neq0$.

\subsubsection{Second FRW Metric Solution}
In this case, by assuming $V_0 \neq 0$ and
\bea
4\ddot{d} - 4\dot{d}^2 -(D-1) H^2 &=& z_0,
\nn\\
2\ddot{d} -(D-1) H^2 &=& 0,
\nn\\
\dot{H} -2H\dot{d}  &=& 0.
\eea
We have
\bea
4d^{\prime\prime} - 4d^{\prime2} -(D-1)\tilde{H}^2 &=& -(V_0 + z_0) ,
\nn\\
2d^{\prime\prime} -(D-1)\tilde{H}^2 &=& 0 ,
\nn\\
\tilde{H}^{\prime} -2\tilde{H}d^{\prime} &=& 0,
\eea
where $ z_0$ is constant.

\begin{figure}
\begin{minipage}[htbp]{0.5\textwidth}
\vspace{5pt}
\scalebox{1.2}[1.2]{\includegraphics[scale=0.32,angle=0]{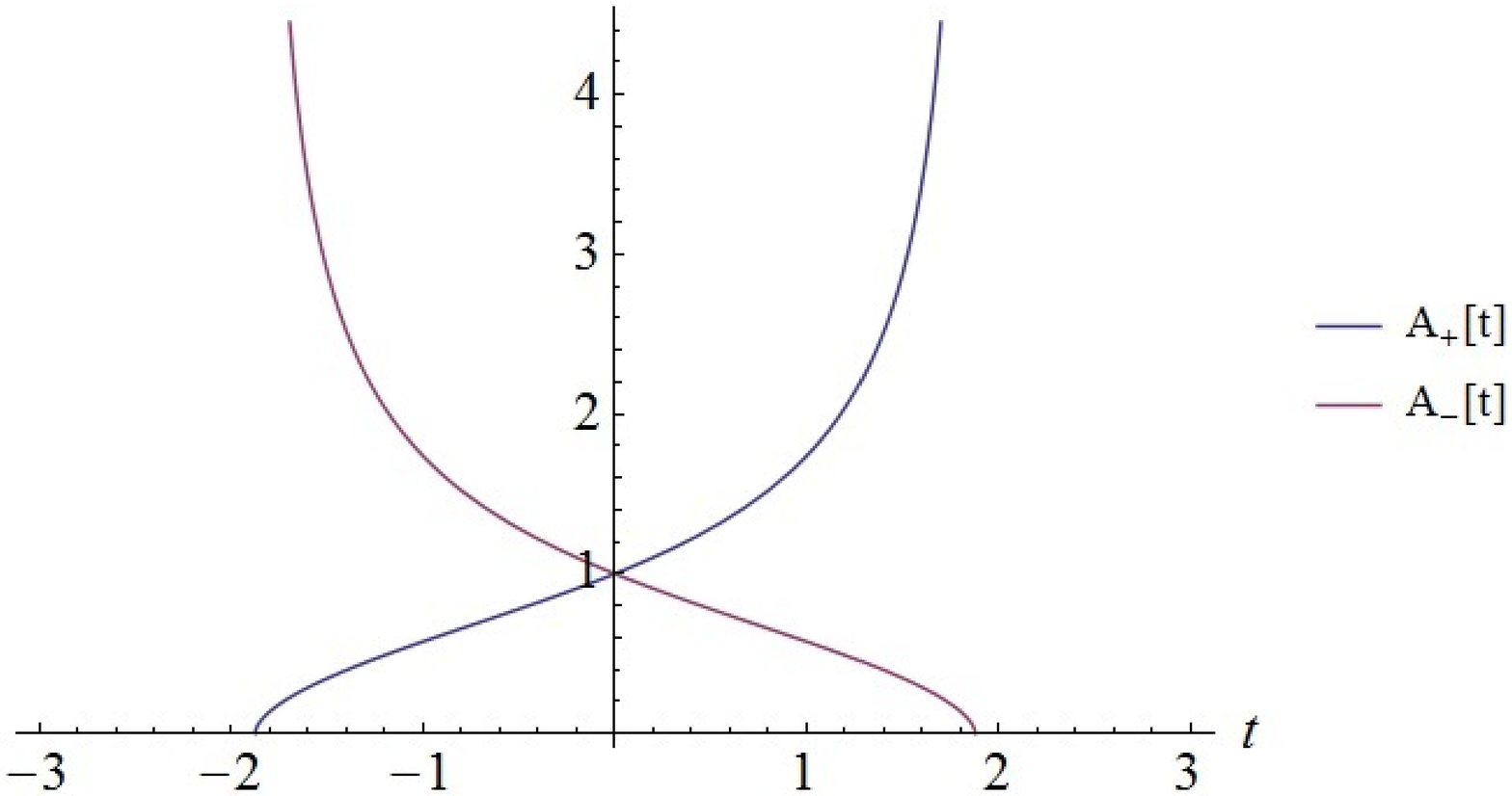}}
\caption{ Two solutions of $A$($t$) in this figure. We can see symmetry of $A$($t$) with respect to origin point ($t=0$). It is possible for behavior of such solution to describe inflationary cosmology. \\ ($D=4$, $z_0=0.7$, $t_0=0$)  }
\label{fig1}
\end{minipage}
\hspace{0.0\textwidth}
\begin{minipage}[htbp]{0.5\textwidth}
\scalebox{1.2}[1.2]{\includegraphics[scale=0.41,angle=0]{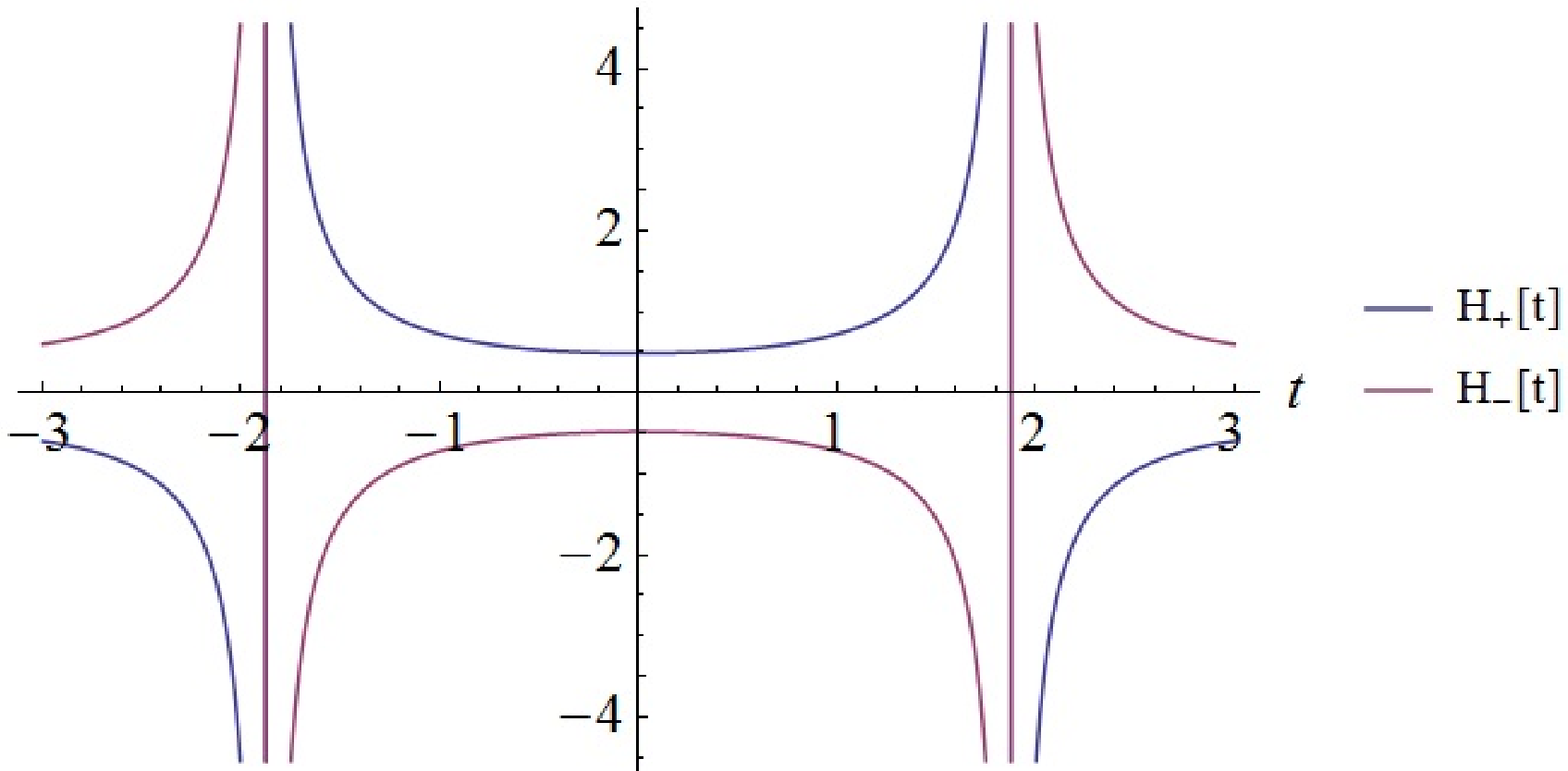}}
\caption{Hubble parameter versus time ($H$$\equiv$$\frac{\dot{a}}{a}$.). \\ ($D=4$, $z_0=0.7$, $t_0=0$)  }
\label{fig2}
\end{minipage}
\end{figure}

Then the solutions of the EOM above are
\bea
a(t, \tilde{t} ) &=& a_0 A(t)\cdot\tilde{A}(\tilde{t}) ,
\nn\\
d(t, \tilde{t} ) &=& -\frac{1}{2}\ln\big[\cos\bigg(\sqrt{z_0}(t+t_0)\bigg)\cos\bigg(\sqrt{-(V_0 + z_0)}(\tilde{t}+\tilde{t}_0)\bigg)\big] + d_0,
\eea
where
\bea
A(t) &=& \exp\bigg[ \frac{2}{\sqrt{D-1}} \tanh^{-1}\big[ \sec(\frac{\sqrt{z_0}}{2}t)\sin\bigg(\frac{\sqrt{z_0}}{2}(t+t_0)\bigg) \big] \bigg]
\nn\\
&& \mbox{or}
\nn\\
&& \exp\bigg[ -\frac{2}{\sqrt{D-1}} \tanh^{-1}\big[ \sec(\frac{\sqrt{z_0}}{2}t)\sin\bigg(\frac{\sqrt{z_0}}{2}(t+t_0)\bigg) \big] \bigg],
\nn\\
\tilde{A}(\tilde{t}) &=& \exp\bigg[ \frac{2}{\sqrt{D-1}} \tanh^{-1}\big[ \sec(\frac{\sqrt{-(V_0 + z_0)}}{2}\tilde{t})\sin\bigg(\frac{\sqrt{-(V_0 + z_0)}}{2}(\tilde{t}+\tilde{t}_0)\bigg) \big] \bigg]
\nn\\
&& \mbox{or}
\nn\\
&& \exp\bigg[ - \frac{2}{\sqrt{D-1}} \tanh^{-1}\big[ \sec(\frac{\sqrt{-(V_0 + z_0)}}{2}\tilde{t})\sin\bigg(\frac{\sqrt{-(V_0 + z_0)}}{2}(\tilde{t}+\tilde{t}_0)\bigg) \big] \bigg],
\nn\\
\eea
and $a_0, d_0, t_0$ are constants.
In this case, $a(t, \tilde{t} ) = a_0 A(t)\cdot\tilde{A}(\tilde{t})$ with a periodic function $\tilde{A}(\tilde{t})$. The period of $\tilde{A}(\tilde{t})$ is $\frac{4\pi}{\sqrt{-(V_0 + z_0)}}$ and $A(t)$ is $\frac{4\pi}{\sqrt{z_0}}$.
Therefore, we can tune the period of $\tilde{A}$ and $A(t)$ by varying the cosmological constant $V_0$ and $z_0$.

If we compactify the dual time coordinate $\tilde{t}$, this solution can potentially describe inflationary cosmology. However $\tilde{A}(\tilde{t})$ is periodic, but not bounded. Therefore it might encounter singularity. Nevertheless, we can expect that the singularity might be smoothed by higher order effects and then $\tilde{A}(\tilde{t})$ will become bounded. \emph{Double field theory only captures the feature of cubic order of closed string field theory}.

By observing the behavior of the function $A(t)$ in Fig.(\ref{fig1}), we can expect that this solution has potential to describe inflationary cosmology. We also check the behavior by calculating the series expansion of it directly,
\bea
&&\tanh^{-1}\big[ \sec(\frac{\sqrt{-V_0}}{2}t)\sin\bigg(\frac{\sqrt{-V_0}}{2}(t+t_0)\bigg) \big]
\nn\\
&=&\frac{1}{2}(t+2t_0)(-V_0)^{\frac{1}{2}} +( \frac{1}{6}t_0^3 + \frac{1}{4}t_0^2 t + \frac{1}{4}t_0 t^2 + \frac{1}{12}t^3 )(-V_0)^{\frac{3}{2}} + O((-V_0)^{\frac{5}{2}}).
\eea
It is easy to find that the leading order contribution of $A(t)$ is indeed the exponential expansion, which is the key feature responsible for solving flatness and horizon problems in many inflation models. We plot two solutions of $A(t)$ and Hubble parameter versus time in Fig.(\ref{fig1}) and Fig.(\ref{fig2}) for $D=4$, $z_0=0.7$ and $t_0=0$.

\subsubsection{Third FRW Metric Solution}
For the third FRW case, let $V_0 \neq 0$ and

\bea
4\ddot{d} - 4\dot{d}^2 -(D-1) H^2 &=& -\frac{V_0}{2} ,
\nn\\
2\ddot{d} -(D-1) H^2 &=& -\frac{V_0}{4} ,
\nn\\
\dot{H} -2H\dot{d}  &=& \pm \frac{V_0}{4\sqrt{D-1}}.
\eea
This means
\bea
4d^{\prime\prime} - 4d^{\prime2} -(D-1)\tilde{H}^2 &=& -\frac{V_0}{2},
\nn\\
2d^{\prime\prime} -(D-1)\tilde{H}^2 &=& -\frac{V_0}{4},
\nn\\
\tilde{H}^{\prime} -2\tilde{H}d^{\prime} &=& \mp \frac{V_0}{4\sqrt{D-1}}.
\eea
The solutions of the EOM now are

\bea
a(t, \tilde{t} ) = a_+ A_+(t)\tilde{A}_+(\tilde{t})\qquad \mbox{or}\qquad a_- A_-(t)\tilde{A}_-(\tilde{t}),
\eea
\bea
d(t, \tilde{t} ) &=& -\frac{1}{2}\ln\big[\cos\bigg(\frac{\sqrt{-V_0}}{2}(t+t_0)\bigg)\cos\bigg(\frac{\sqrt{-V_0}}{2}(\tilde{t}+\tilde{t}_0)\bigg)\big]  + d_0,
\eea
where
\bea
A_\pm(t) = \bigg[ \cos\bigg(\frac{\sqrt{-V_0}}{2}(t+t_0)\bigg) \bigg]^{\pm \frac{1}{\sqrt{D-1}}},
\qquad
\tilde{A}_\pm(\tilde{t}) = \bigg[ \cos\bigg(\frac{\sqrt{-V_0}}{2}(\tilde{t}+\tilde{t}_0)\bigg) \bigg]^{\pm \frac{1}{\sqrt{D-1}}},
\eea
and $a_\pm$, $d_0$, $t_0$ are constants.
In this case, $A(t)$ and $\tilde{A}(\tilde{t})$ are just sinusoidal functions ($V_0<0$) or hyperbolic functions ($V_0>0$).

When $V_0<0$, $A(t)$ and $\tilde{A}(\tilde{t})$ will be imaginary. This solution cannot have real physical meaning.

On the other hand, for the case $V_0>0$, $A(t)$ and $\tilde{A}(\tilde{t})$ are hyperbolic functions and thus not periodic. We cannot use it to describe our universe. We plot two solutions of $A(t)$ and Hubble parameter versus time in Fig.(\ref{fig1}) and Fig.(\ref{fig2}) for $D=4$, $V_0=2$ and $t_0=0$.

\begin{figure}
\begin{minipage}[h]{0.5\textwidth}
\vspace{5pt}
\scalebox{1.2}[1.2]{\includegraphics[scale=0.32,angle=0]{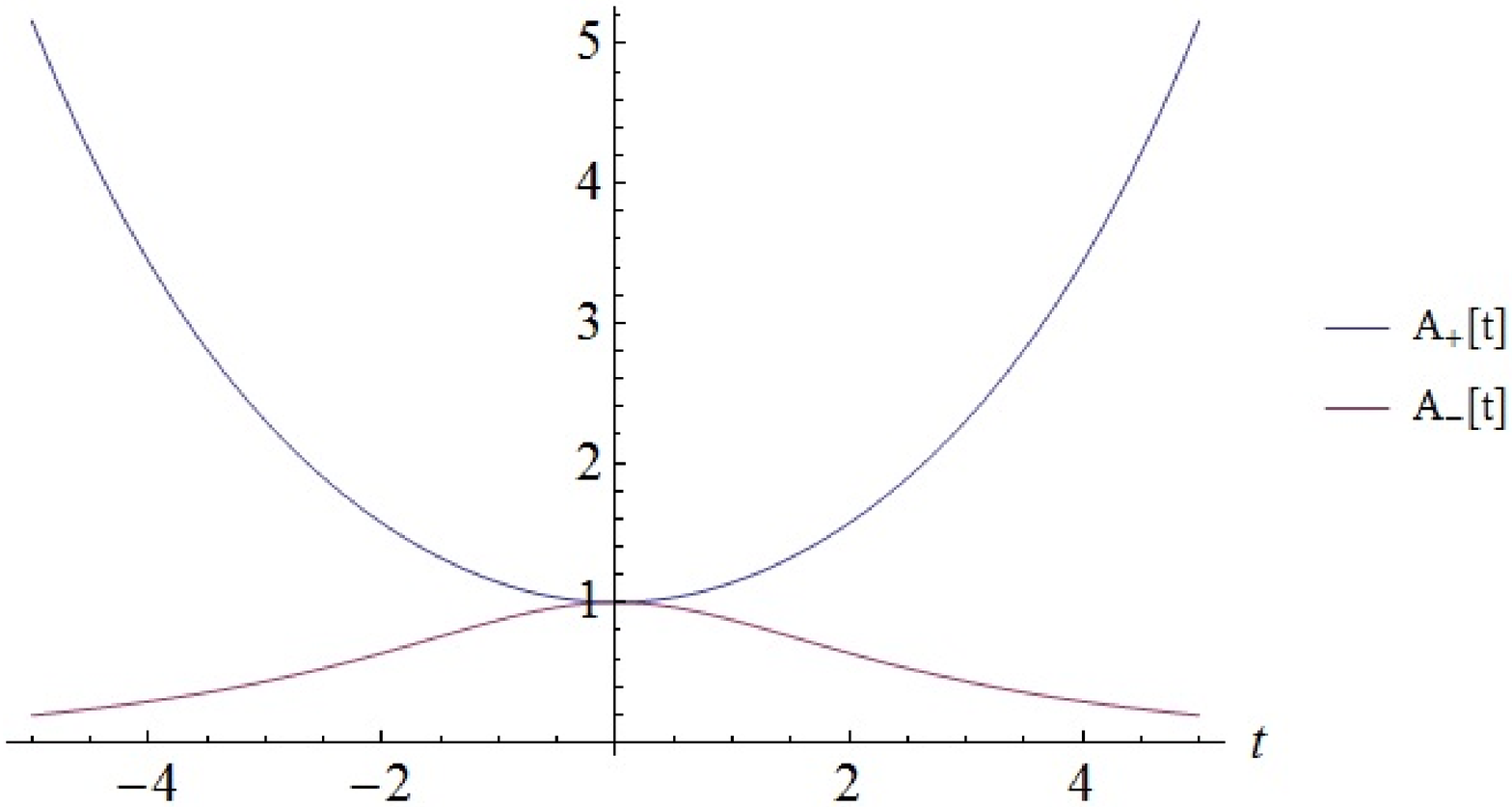}}
\caption{Two solutions for $A$($t$) for $V_0>0$. \\ ($D=4$, $V_0=2$, $t_0=0$) }
\label{fig3}
\end{minipage}
\hspace{0.0\textwidth}
\begin{minipage}[h]{0.5\textwidth}
\scalebox{1.2}[1.2]{\includegraphics[scale=0.32,angle=0]{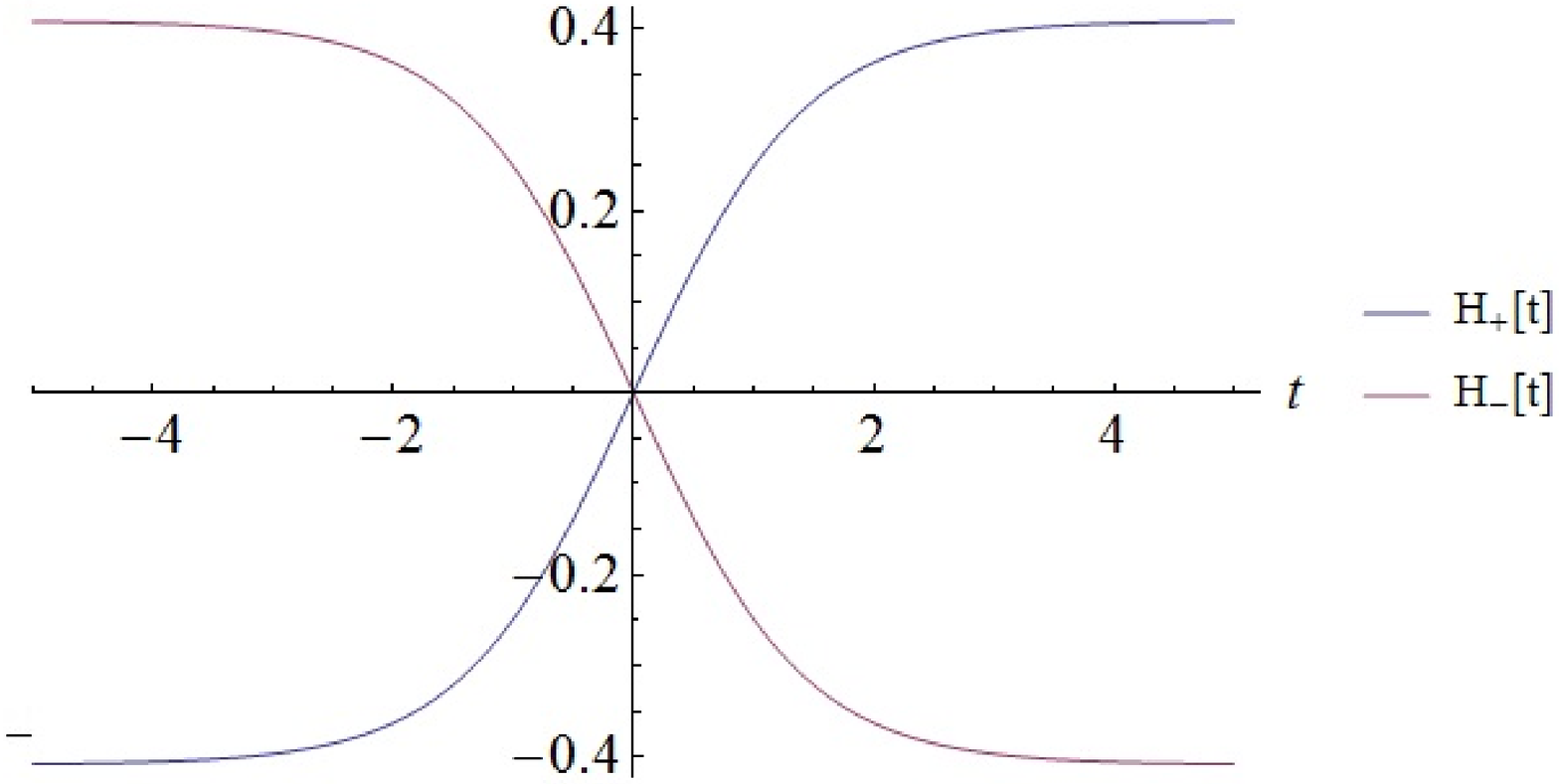}}
\caption{Hubble parameter versus time for $V_0>0$. ($D=4$, $V_0=2$, $t_0=0$) }
\label{fig4}
\end{minipage}
\end{figure}

\subsection{Bianchi Type Spacetime and Stability Issue}
In the following, we investigate the stability issue by using the Bianchi type metric field $g_{ij} = \mbox{diag} \bigg(-1, a_1(t), a_2(t), \cdots, a_2(t) \bigg)$, $b_{ij} = 0$, and then the EOM are
\bea
{\cal R}&=&0 \Rightarrow \bigg(4d^{\prime\prime} - 4d^{\prime2} -\tilde{H}_1^2 -(D-2)\tilde{H}_2^2\bigg) + \bigg(4\ddot{d} - 4\dot{d}^2 -H_1^2 -(D-2)H_2^2\bigg) + V_0 = 0,
\nn\\
{\cal R}_{tt}&=&0 \Rightarrow \bigg( 2d^{\prime\prime} -\tilde{H}_1^2 -(D-2)\tilde{H}_2^2 \bigg) - \bigg( 2\ddot{d} -H_1^2 -(D-2)H_2^2 \bigg) = 0,
\nn\\
{\cal R}_{xx}&=&0 \Rightarrow ( \tilde{H_1}^{\prime} -2\tilde{H_1}d' ) + ( \dot{H_1} -2H_1\dot{d} )  = 0,
\nn\\
{\cal R}_{ii}&=&0 \Rightarrow ( \tilde{H_2}^{\prime} -2\tilde{H_2}d' ) + ( \dot{H_2} -2H_2\dot{d} )  = 0,
\eea
where $i\neq x$ and $t$, and $H_1 \equiv \frac{\dot{a_1}}{a_1}$, $\tilde{H}_1 \equiv \frac{a_1^{\prime}}{a_1}$, $H_2 \equiv \frac{\dot{a_2}}{a_2}$, $\tilde{H}_2 \equiv \frac{a_2^{\prime}}{a_2}$.

\subsubsection{First Solution}
Similar to the first FRW metric solution, we consider
\bea
4\ddot{d} - 4\dot{d}^2 -H_1^2 -(D-2)H_2^2 &=& 0,
\nn\\
2\ddot{d} -H_1^2 -(D-2)H_2^2 &=& 0,
\nn\\
\dot{H_1} -2H_1\dot{d}  &=& 0,
\nn\\
\dot{H_2} -2H_2\dot{d}  &=& 0.
\eea
These imply
\bea
4d^{\prime\prime} - 4d^{\prime2} -\tilde{H_1}^2 -(D-2)\tilde{H_2}^2 &=& 0,
\nn\\
2d^{\prime\prime} -\tilde{H_1}^2 -(D-2)\tilde{H_2}^2 &=& 0,
\nn\\
\tilde{H_1}^{\prime} -2\tilde{H_1}d^{\prime} &=& 0,
\nn\\
\tilde{H_2}^{\prime}-2\tilde{H_2}d^{\prime} &=& 0,
\eea
and then the solutions of EOM are
\bea
a_1(t, \tilde{t} ) = a_0 A_1(t)\cdot\tilde{A_1}(\tilde{t}) ,
\qquad
a_2(t, \tilde{t} ) = a_0 A_2(t)\cdot\tilde{A_2}(\tilde{t}) ,
\eea
\bea
d(t, \tilde{t} ) = -\frac{1}{2} \ln\bigg((t+t_0)(\tilde{t}+\tilde{t}_0) \bigg) + d_0,
\eea
where
\bea
A_1(t)&=&B(t)\cdot B_\pm(t),\ A_2(t)=\frac{B(t)}{ B_\pm(t)},\ \tilde{A_1}(\tilde{t})=\tilde{B}(t)\cdot \tilde{B}_\pm(t),\ \tilde{A_2}(\tilde{t})=\frac{\tilde{B}(t)}{ \tilde{B}_\pm(t)},
\nn\\
B(t)&=& (t+t_0)^{p},\ B_\pm(t)= (t+t_0)^{p_\pm},\ \tilde{B}(\tilde{t})= (\tilde{t}+\tilde{t}_0)^{\tilde{p}},\ \tilde{B}_\pm(\tilde{t})=(\tilde{t}+\tilde{t}_0)^{\tilde{p}_\pm},
\nn\\
p_\pm &=& \frac{(D-3)p\pm \sqrt{(D-1)-4(D-2)p^2}}{D-1},\ \tilde{p}_\pm = \frac{(D-3)\tilde{p}\pm \sqrt{(D-1)-4(D-2)\tilde{p}^2}}{D-1},
\nn
\eea
in which $(t+t_0)(\tilde{t}+\tilde{t}_0)$$\neq$ $0$ and $t_0$, $\tilde{t}_0$, $a_0$, $d_0$, $p$, $\tilde{p}$ are constants.

If we compactify the dual time coordinate, we will obtain the following EOM only:
\bea
{\cal R}&=&0 \Rightarrow 4\ddot{d} - 4\dot{d}^2 -H_1^2 -(D-2)H_2^2 = 0,
\nn\\
{\cal R}_{tt}&=&0 \Rightarrow 2\ddot{d} -H_1^2 -(D-2)H_2^2 = 0,
\nn\\
{\cal R}_{xx}&=&0 \Rightarrow \dot{H_1} -2H_1\dot{d} = 0,
\nn\\
{\cal R}_{ii}&=&0 \Rightarrow \dot{H_2} -2H_2\dot{d} = 0,
\eea
where $i\neq x$ and $t$, and $H_1 \equiv \frac{\dot{a_1}}{a_1}$, $H_2 \equiv \frac{\dot{a_2}}{a_2}$.
Then the solutions of the above EOM are
\bea
a_1(t) =B_1(t) \cdot B_2(t),\qquad a_2(t) = \frac{B_1(t)}{B_2(t)},
\eea
\bea
d(t) = -\frac{1}{2}\bigg( \ln(t+t_0) \bigg) + d_0,
\eea
where
\bea
B_1(t)= a_1 (t+t_0)^{p_1},\qquad B_2(t)= a_2 (t+t_0)^{p_2}.
\eea
\bea
p_2 = \frac{(D-3)p_1\pm \sqrt{(D-1)-4(D-2)p_1^2}}{D-1}
\eea
and $p_1$ is an arbitrary constant.

In order to use this solution to describe our universe, we will refer to $B_1(t)$ as background part with $p_1 > 0$ and refer to $B_2(t)$ as non-isotropic perturbation.
Therefore $p_2$ contains at least one positive solution when assumeing $D\ge3$ and then in the late time $B_2(t)$ will be much larger than $1$ in general. Therefore the solution is unstable to non-isotropic perturbation.

\subsubsection{Second Solution}
Similar to the second FRW metric solution, let $V_0 \neq 0$ and
\bea
4\ddot{d} - 4\dot{d}^2 -H_1^2 -(D-2)H_2^2 &=& z_0,
\nn\\
2\ddot{d} -H_1^2 -(D-2)H_2^2 &=& 0,
\nn\\
\dot{H_1} -2H_1\dot{d}  &=& 0,
\nn\\
\dot{H_2} -2H_2\dot{d}  &=& 0,
\eea
which leads to
\bea
4d^{\prime\prime} - 4d^{\prime 2} -\tilde{H_1}^2 -(D-2)\tilde{H_2}^2 &=& -(V_0+z_0),
\nn\\
2d^{\prime\prime} -\tilde{H_1}^2 -(D-2)\tilde{H_2}^2 &=& 0,
\nn\\
\tilde{H_1}^{\prime} -2\tilde{H_1}d^{\prime} &=& 0,
\nn\\
\tilde{H_2}^{\prime} -2\tilde{H_2}d^{\prime} &=& 0.
\eea
The solutions of EOM are
\bea
a_1(t, \tilde{t} ) = a_0 A_1(t)\cdot\tilde{A_1}(\tilde{t}) ,\qquad a_2(t, \tilde{t} ) = a_0 A_2(t)\cdot\tilde{A_2}(\tilde{t}) ,
\eea
\bea
d(t, \tilde{t} ) = -\frac{1}{2} \ln\bigg((t+t_0)(\tilde{t}+\tilde{t_0}) \bigg) + d_0,
\eea
where
\bea
A_1(t)&=&B(t)\cdot B_\pm(t),\
A_2(t)= \frac{B(t)}{B_\pm(t)},\
\tilde{A_1}(\tilde{t})=\tilde{B}(t)\cdot \tilde{B}_\pm(t),\
\tilde{A_2}(\tilde{t})= \frac{\tilde{B}(t)}{\tilde{B}_\pm(t)},
\nn\\
B(t)&=& \exp\bigg[ \frac{2p}{\sqrt{z_0}} \tanh^{-1}\big[ \sec(\frac{\sqrt{z_0}}{2}t)\sin\bigg(\frac{\sqrt{z_0}}{2}(t+t_0)\bigg) \big] \bigg],
\nn\\
B_\pm(t)&=& \exp\bigg[ \frac{2p_\pm}{\sqrt{z_0}} \tanh^{-1}\big[ \sec(\frac{\sqrt{z_0}}{2}t)\sin\bigg(\frac{\sqrt{z_0}}{2}(t+t_0)\bigg) \big] \bigg],
\nn\\
\tilde{B}(\tilde{t})&=& \exp\bigg[ \frac{2\tilde{p}}{\sqrt{-(V_0+z_0)}} \tanh^{-1}\big[ \sec(\frac{\sqrt{-(V_0+z_0)}}{2}\tilde{t})\sin\bigg(\frac{\sqrt{-(V_0+z_0)}}{2}(\tilde{t}+\tilde{t_0})\bigg) \big] \bigg],
\nn\\
\tilde{B}_\pm(\tilde{t})&=&\exp\bigg[ \frac{2\tilde{p_\pm}}{\sqrt{-(V_0+z_0)}} \tanh^{-1}\big[ \sec(\frac{\sqrt{-(V_0+z_0)}}{2}\tilde{t})\sin\bigg(\frac{\sqrt{-(V_0+z_0)}}{2}(\tilde{t}+\tilde{t_0})\bigg) \big] \bigg],
\nn\\
p_\pm &=& \frac{(D-3)p\pm \sqrt{(D-1)z_0-4(D-2)p^2}}{D-1},
\nn\\
\tilde{p}_\pm &=& \frac{(D-3)\tilde{p}\pm \sqrt{-(D-1)(V_0+z_0)-4(D-2)\tilde{p}^2}}{D-1},
\nn
\eea
and $t_0$, $\tilde{t_0}$, $a_0$, $d_0$, $p$, $\tilde{p}$ are constants.
As in previous section, we expect that this solution can be compactified, and then similar to the first case, we will take $B(t)$ as background part and $B_\pm(t)$ as non-isotropic perturbations. This solution is also unstable to non-isotropic perturbation.

\subsubsection{Third Solution}
Similar to the third FRW metric solution, by assuming $V_0 \neq 0$ and
\bea
4\ddot{d} - 4\dot{d}^2 -H_1^2 -(D-2)H_2^2 &=& -\frac{V_0}{2},
\nn\\
2\ddot{d} -H_1^2 -(D-2)H_2^2 &=& -\frac{V_0}{4},
\nn\\
\dot{H_1} -2H_1\dot{d}  &=& p+p_\pm,
\nn\\
\dot{H_2} -2H_2\dot{d}  &=& p-p_\pm,
\eea
we obtain
\bea
4d^{\prime\prime} - 4d^{\prime 2} -\tilde{H_1}^2 -(D-2)\tilde{H_2}^2 &=& -\frac{V_0}{2},
\nn\\
2d^{\prime\prime} -\tilde{H_1}^2 -(D-2)\tilde{H_2}^2 &=& -\frac{V_0}{4},
\nn\\
\tilde{H_1}^{\prime} -2\tilde{H_1}d^{\prime} &=& -(p+p_\pm),
\nn\\
\tilde{H_2}^{\prime} -2\tilde{H_2}d^{\prime} &=& -(p-p_\pm).
\eea
The solutions of EOM are
\bea
a_1(t, \tilde{t} ) = a_0 A_1(t)\cdot\tilde{A_1}(\tilde{t}) ,\qquad
a_2(t, \tilde{t} ) = a_0 A_2(t)\cdot\tilde{A_2}(\tilde{t}) ,
\eea
\bea
d(t, \tilde{t} ) = -\frac{1}{2} \ln\bigg((t+t_0)(\tilde{t}+\tilde{t_0}) \bigg) + d_0,
\eea
where
\bea
A_1(t)&=&B(t)\cdot B_\pm(t),\
A_2(t)= \frac{B(t)}{B_\pm(t)},\
\tilde{A_1}(\tilde{t})=\tilde{B}(t)\cdot \tilde{B}_\pm(t),\
\tilde{A_2}(\tilde{t})= \frac{\tilde{B}(t)}{\tilde{B}_\pm(t)},
\nn\\
B(t)&=& \bigg[ \cos\bigg(\frac{\sqrt{-V_0}}{2}(t+t_0)\bigg) \bigg]^{\frac{4p}{V_0}},\
B_\pm(t)= \bigg[ \cos\bigg(\frac{\sqrt{-V_0}}{2}(t+t_0)\bigg) \bigg]^{\frac{4p_\pm}{V_0}},
\nn\\
\tilde{B}(\tilde{t})&=& \bigg[ \cos\bigg(\frac{\sqrt{-V_0}}{2}(\tilde{t}+\tilde{t_0})\bigg) \bigg]^{\frac{-4p}{V_0}},\
\tilde{B}_\pm(\tilde{t})= \bigg[ \cos\bigg(\frac{\sqrt{-V_0}}{2}(\tilde{t}+\tilde{t_0})\bigg) \bigg]^{\frac{-4p_\pm}{V_0}},
\nn\\
p_\pm &=& \frac{(D-3)p\pm \sqrt{\frac{(D-1)z_0^2}{4}-4(D-2)p^2}}{D-1},
\nn
\eea
and $t_0$, $\tilde{t_0}$, $a_0$, $d_0$, $p$ are constants. This solution is not useful to describe our universe as the same discussion in the previous section.

\subsection{Doubled Spherically Symmetric Metric}
 In the following we show that our spacetime depends only on $r$ and $\tilde{r}$. We expect that we can find some solutions which are analogous to spherically symmetric metric. Now, based on $O(D,D)$ or manifest T-duality, we expect three dimensional metric field of the form $g_{\mu\nu} = \mbox{diag} \bigg( -f(r,\tilde{r}), h(r,\tilde{r}), (\frac{r}{\tilde{r}})^2 \bigg)$ which we called \lq\lq doubled spherically symmetric metric\rq\rq in the following discussion.

By assuming $g_{\mu\nu} = \mbox{diag} \bigg( -f(r,\tilde{r}), h(r,\tilde{r}), (\frac{r}{\tilde{r}})^2 \bigg)$, $b_{ij} = 0$, and $V_0=0$, the EOM are
\bea
{\cal R}&=&0 \Rightarrow h \bigg( \frac{1}{4} \tilde{X}^2 + \frac{3}{4} \tilde{Z}^2 + 4 \tilde{Y}^2 + 4 \tilde{Y}\tilde{Z} - (\tilde{Z}^{\prime} + 4 \tilde{Y}^{\prime} ) + \tilde{r}^{-2} \bigg)
\nn\\
&&\ \ \ \ \ \ + h^{-1}\bigg( \frac{1}{4} X^2 + \frac{3}{4} Z^2 + 4 Y^2 + 4 YZ - (\dot{Z} + 4 \dot{Y} ) + r^{-2} \bigg) = 0,
\nn\\
{\cal R}_{tt}&=&0 \Rightarrow h\bigg( \tilde{X}\tilde{Z}+\tilde{X}^{\prime}-2\tilde{Y}\tilde{X} \bigg)+ h^{-1}\bigg( -XZ+\dot{X}-2YX \bigg) = 0,
\nn\\
{\cal R}_{rr}&=&0 \Rightarrow h\bigg( -\tilde{X}^2-\tilde{Z}^2+4\tilde{Y}\tilde{Z}-2\tilde{Z}^{\prime}+8\tilde{Y}^{\prime}
-4\tilde{r}^{-2} \bigg)
\nn\\
&&\ \ \ \ \ \ - h^{-1}\bigg( -X^2-Z^2-4YZ+2\dot{Z}+8\dot{Y}-4r^{-2} \bigg) = 0,
\nn\\
{\cal R}_{\theta\theta}&=&0 \Rightarrow  \frac{h}{\tilde{r}}(-\tilde{Z}+2\tilde{Y}+\tilde{r}^{-1}) - \frac{h^{-1}}{r} (Z+2Y+r^{-1}) = 0,
\eea
where
\bea
X &\equiv& \frac{\dot{f}}{f},\qquad Y \equiv \dot{d},\qquad Z \equiv \frac{\dot{h}}{h},
\nn\\
\tilde{X} &\equiv& \frac{f^{\prime}}{f},\ \ \ \ \ \tilde{Y} \equiv d^{\prime},\ \ \ \ \ \tilde{Z} \equiv \frac{h^{\prime}}{h}.
\eea
We denote ${}^{\prime} \equiv \partial_{\tilde{r}}$ and $ \dot{} \equiv \partial_{r}$ when discussing doubled spherically symmetric metric.

Assume
\bea
\frac{1}{4} X^2 + \frac{3}{4} Z^2 + 4 Y^2 + 4 YZ - (\dot{Z} + 4 \dot{Y} ) + r^{-2} &=& 0,
\nn\\
-XZ+\dot{X}-2YX &=& 0,
\nn\\
-X^2-Z^2-4YZ+2\dot{Z}+8\dot{Y}-4r^{-2} &=& 0,
\nn\\
Z+2Y+r^{-1} &=& 0.
\eea
These assumptions imply
\bea
\frac{1}{4} \tilde{X}^2 + \frac{3}{4} \tilde{Z}^2 + 4 \tilde{Y}^2 + 4 \tilde{Y}\tilde{Z} - (\tilde{Z}' + 4 \tilde{Y}' ) + \tilde{r}^{-2} &=& 0,
\nn\\
\tilde{X}\tilde{Z}+\tilde{X}'-2\tilde{Y}\tilde{X} &=& 0,
\nn\\
 -\tilde{X}^2-\tilde{Z}^2+4\tilde{Y}\tilde{Z}-2\tilde{Z}'+8\tilde{Y}'-4\tilde{r}^{-2} &=& 0,
\nn\\
-\tilde{Z}+2\tilde{Y}+\tilde{r}^{-1} &=& 0.
\eea
The solutions are
\bea
f(r,\tilde{r}) &=& f_0 r^{x_0} \tilde{r}^{x_1},
\nn\\
h(r,\tilde{r}) &=& h_0 r^{-(1+2y_0)} \tilde{r}^{1+2y_1},
\nn\\
d(r,\tilde{r}) &=& y_0 \ln r + y_1 \ln \tilde{r} + d_0,
\eea
where $x_0^2 - (2y_0+1)(2y_0-3) = 0$, $x_1^2 - (2y_1+1)(2y_1-3) = 0$ and $f_0$, $h_0$, $x_0$, $x_1$, $y_0$, $y_1$ are constants. If we choose $x_0=x_1=0$, we can compactify one dual time to study one time physics in this solution. We found the equations of motion of metric field $g_{\mu\nu}$=$\mbox{diag}\bigg(-f(r)$, $h(r)$, $r^2\bigg)$ is the same as the solution of equations of motion in ordinary coordinates. We can also observe $X\leftrightarrow X$, $Y\leftrightarrow Y$ and $Z\leftrightarrow -Z$ by exchanging $r$ and $\tilde{r}$. According to this observation we can find solutions readily. The solution of dual coordinate is just mapped from the solution in ordinary coordinates by the manifest T-duality. It is also an analogue of spherically symmetric solution based on $O( D, D)$. That is why we call this solution doubled spherically symmetric metric. If we want to find analogue solutions to general relativity based on $O(D, D)$, this example shows that we cannot just use the same metric as general relativity.

\section{Discussion and Conclusion}
\label{4}
In this paper, we show that solutions of FRW and doubled spherically symmetric metric in DFT of massless closed string field theory, and then discuss the stability property of FRW metric.

We found some possible cosmological model in FRW metric solutions in double field theory. We found our solution have the potential to become an inflationary model if we compactify the dual time coordinate. We also performed test on anisotropic perturbation. The conclusion is that those FRW solutions are unstable to anisotropic perturbation, therefore there is a fine-tuning problem if we want to apply the FRW solution to the standard cosmology.
Furthermore, we also found solutions of doubled spherically symmetric metric. This solution of ordinary coordinates correspond to spherical symmetry. The dual coordinate of the solution can be mapped by manifest T-duality from the solution of the ordinary coordinates.  We also expect deeper understanding of classical solutions can shed light on the relax of constraint and  Scherk-Schwarz compactification. Relation of constraint is a very important problem in double field theory. From the calculation of four tachyon amplitude, we already knew that strong constraint must be imposed in all the non-compact directions, but the compact directions are not restricted by any constraints \cite{Betz:2014aia}. It is consistent with our current conjecture, since we only require weak constraints in the compact directions. Because some fluxes do not appear when we perform compactification, the constraint is expected to be relaxed.

It is also interesting to formulate the low energy brane theory \cite{Ho:2013iia} with background effect \cite{Lawrence:2006ma} in double field theory, an issue that is still poorly understood. Recently, some people already showed how to combine non-commutative geometry with generalized geometry \cite{Asakawa:2012px}. Extending D-brane theory to doubled geometry could be an interesting problem. We expect that brane cosmology of double field theory could provide some insights in this direction. Another interesting direction would be to understand more about the entropy of black hole \cite{Berman:2011kg}. We leave these interesting explorations as future works.

Finally, we give some comments on relaxing constraint. We find that the solutions on doubled space can satisfy neither the weak nor the strong constraint.  But we can use suitable field redefinitions to let our solutions satisfy the weak constraint. For example, Hubble parameters ($H$, $\tilde{H}$) satisfy the weak constraint, but the metric does not. Even if we require weak constraint only, the solutions of DFT are already restricted. If we do not use any field redefinition, many solutions of double field theory do not satisfy the weak constraint. By using field redefinitions, we can have non-trivial solutions with weak constraint. Technically, we can use field redefinitions to make non-trivial solutions satisfy weak constraint, and then check that $\partial\tilde{\partial}\delta (\cdots)$. Solving equations of motion also give us more hints about how to perform field redefinitions to achieve the goal of constraint relaxation.

\section*{Acknowledgement}
We thank David S. Berman, Pei-Ming Ho, Olaf Hohm, Keisuke Izumi, Kanghoon Lee, Yen Chin Ong, Jeong-Hyuck Park and Yoonji Suh
for useful discussions. Chen-Te Ma wants to especially to thank Hadi Godazgar, Imtak Jeon and Pei-Wen Peggy Kao for their nice suggestions.
This work is supported in part by
NTU (grant \#NTU-CDP-102R7708), National Science Council, and by Leung Center for Cosmology and
Particle Astrophysics, Taiwan, R.O.C.


\vskip .8cm
\baselineskip 22pt
\begin{thebibliography}{99}
\itemsep 0pt

\bibitem{Hull:2009mi}
  C.~Hull and B.~Zwiebach,
  ``Double Field Theory,''  JHEP {\bf 0909}, 099 (2009)  [arXiv:0904.4664 [hep-th]].  

\bibitem{Hull:2009zb}
  C.~Hull and B.~Zwiebach,
  ``The Gauge algebra of double field theory and Courant brackets,''  JHEP {\bf 0909}, 090 (2009)  [arXiv:0908.1792 [hep-th]].  

\bibitem{Hohm:2010jy}
  O.~Hohm, C.~Hull and B.~Zwiebach,
  ``Background independent action for double field theory,''  JHEP {\bf 1007}, 016 (2010)  [arXiv:1003.5027 [hep-th]].  

\bibitem{Hohm:2010pp}
  O.~Hohm, C.~Hull and B.~Zwiebach,
  ``Generalized metric formulation of double field theory,''  JHEP {\bf 1008}, 008 (2010)  [arXiv:1006.4823 [hep-th]].  

\bibitem{Siegel:1993xq}
  W.~Siegel,
  ``Two vierbein formalism for string inspired axionic gravity,''
  Phys.\ Rev.\ D {\bf 47}, 5453 (1993)
  [hep-th/9302036].
  W.~Siegel,
  ``Superspace duality in low-energy superstrings,''
  Phys.\ Rev.\ D {\bf 48}, 2826 (1993)
  [hep-th/9305073].
  W.~Siegel,
  ``Manifest duality in low-energy superstrings,''
  In *Berkeley 1993, Proceedings, Strings '93* 353-363, and State U. New York Stony Brook - ITP-SB-93-050 (93,rec.Sep.) 11 p. (315661)
  [hep-th/9308133].

\bibitem{Hellerman:2002ax}
  S.~Hellerman, J.~McGreevy and B.~Williams,
  ``Geometric constructions of nongeometric string theories,''  JHEP {\bf 0401}, 024 (2004)  [hep-th/0208174].  
  A.~Dabholkar and C.~Hull,
  ``Duality twists, orbifolds, and fluxes,''  JHEP {\bf 0309}, 054 (2003)  [hep-th/0210209].  
  A.~Flournoy, B.~Wecht and B.~Williams,
  ``Constructing nongeometric vacua in string theory,''  Nucl.\ Phys.\ B {\bf 706}, 127 (2005)  [hep-th/0404217].  

\bibitem{Hohm:2011zr}
  O.~Hohm, S.~K.~Kwak and B.~Zwiebach,
  ``Unification of Type II Strings and T-duality,''  Phys.\ Rev.\ Lett.\  {\bf 107}, 171603 (2011)  [arXiv:1106.5452 [hep-th]].  
  O.~Hohm, S.~K.~Kwak and B.~Zwiebach,
  ``Double Field Theory of Type II Strings,''  JHEP {\bf 1109}, 013 (2011)  [arXiv:1107.0008 [hep-th]].  
  O.~Hohm and S.~K.~Kwak,
  ``Frame-like Geometry of Double Field Theory,''  J.\ Phys.\ A {\bf 44}, 085404 (2011)  [arXiv:1011.4101 [hep-th]].  
  O.~Hohm and S.~K.~Kwak,
  ``N=1 Supersymmetric Double Field Theory,''  JHEP {\bf 1203}, 080 (2012)  [arXiv:1111.7293 [hep-th]].  
  O.~Hohm and B.~Zwiebach,
  ``On the Riemann Tensor in Double Field Theory,''  JHEP {\bf 1205}, 126 (2012)  [arXiv:1112.5296 [hep-th]].  
  O.~Hohm and B.~Zwiebach,
  ``Towards an invariant geometry of double field theory,''  J.\ Math.\ Phys.\  {\bf 54}, 032303 (2013)  [arXiv:1212.1736 [hep-th]].  
  D.~S.~Berman, C.~D.~A.~Blair, E.~Malek and M.~J.~Perry,
  ``The $O_{D,D}$ Geometry of String Theory,''  arXiv:1303.6727 [hep-th].  
  D.~S.~Berman, H.~Godazgar, M.~Godazgar and M.~J.~Perry,
  ``The Local symmetries of M-theory and their formulation in generalised geometry,''  JHEP {\bf 1201}, 012 (2012)  [arXiv:1110.3930 [hep-th]].  
  J.~-H.~Park,
  ``Comments on double field theory and diffeomorphisms,''  JHEP {\bf 1306}, 098 (2013)  [arXiv:1304.5946 [hep-th]].  
  K.~Lee and J.~-H.~Park,
  ``Covariant action for a string in $\mathit{doubled}$ $\mathit{yet}$ $\mathit{gauged}$ spacetime,''  Nucl.\ Phys.\ B {\bf 880}, 134 (2014)  [arXiv:1307.8377 [hep-th]].  
  I.~Jeon, K.~Lee and J.~-H.~Park,
  ``Differential geometry with a projection: Application to double field theory,''  JHEP {\bf 1104}, 014 (2011)  [arXiv:1011.1324 [hep-th]].  
  I.~Jeon, K.~Lee and J.~-H.~Park,
  ``Stringy differential geometry, beyond Riemann,''  Phys.\ Rev.\ D {\bf 84}, 044022 (2011)  [arXiv:1105.6294 [hep-th]].  
  J.~-H.~Park and Y.~Suh,
  ``U-gravity : ${\mathbf{SL}(N)}$,''  arXiv:1402.5027 [hep-th].  

\bibitem{Andriot:2011uh}
  D.~Andriot, M.~Larfors, D.~Lust and P.~Patalong,
  ``A ten-dimensional action for non-geometric fluxes,''  JHEP {\bf 1109}, 134 (2011)  [arXiv:1106.4015 [hep-th]].  

\bibitem{Andriot:2013xca}
  D.~Andriot and A.~eBetz,
  ``$\beta$-supergravity: a ten-dimensional theory with non-geometric fluxes, and its geometric framework,''  JHEP {\bf 1312}, 083 (2013)  [arXiv:1306.4381 [hep-th]].  
  D.~Andriot, O.~Hohm, M.~Larfors, D.~Lust and P.~Patalong,
  ``A geometric action for non-geometric fluxes,''  Phys.\ Rev.\ Lett.\  {\bf 108}, 261602 (2012)  [arXiv:1202.3060 [hep-th]].  
  D.~Andriot, O.~Hohm, M.~Larfors, D.~Lust and P.~Patalong,
  ``Non-Geometric Fluxes in Supergravity and Double Field Theory,''  Fortsch.\ Phys.\  {\bf 60}, 1150 (2012)  [arXiv:1204.1979 [hep-th]].  
  R.~Blumenhagen, A.~Deser, E.~Plauschinn and F.~Rennecke,
  ``A bi-invariant Einstein-Hilbert action for the non-geometric string,''  Phys.\ Lett.\ B {\bf 720}, 215 (2013)  [arXiv:1210.1591 [hep-th]].  
  R.~Blumenhagen, A.~Deser, E.~Plauschinn and F.~Rennecke,
  ``Non-geometric strings, symplectic gravity and differential geometry of Lie algebroids,''  JHEP {\bf 1302}, 122 (2013)  [arXiv:1211.0030 [hep-th]].  
  R.~Blumenhagen, A.~Deser, E.~Plauschinn, F.~Rennecke and C.~Schmid,
  ``The Intriguing Structure of Non-geometric Frames in String Theory,''  Fortsch.\ Phys.\  {\bf 61}, 893 (2013)  [arXiv:1304.2784 [hep-th]].  
  M.~Grana, R.~Minasian, M.~Petrini and D.~Waldram,
  ``T-duality, Generalized Geometry and Non-Geometric Backgrounds,''  JHEP {\bf 0904}, 075 (2009)  [arXiv:0807.4527 [hep-th]].  
  G.~Aldazabal, W.~Baron, D.~Marques and C.~Nunez,
  ``The effective action of Double Field Theory,''  JHEP {\bf 1111}, 052 (2011)  [Erratum-ibid.\  {\bf 1111}, 109 (2011)]  [arXiv:1109.0290 [hep-th]].  

\bibitem{Hohm:2013jaa}
  O.~Hohm, W.~Siegel and B.~Zwiebach,
  ``Doubled $\alpha'$-geometry,''  JHEP {\bf 1402}, 065 (2014)  [arXiv:1306.2970 [hep-th]].  

\bibitem{Hohm:2013uia}
  O.~Hohm and H.~Samtleben,
  ``Exceptional Field Theory II: E$_{7(7)}$,''
  Phys.\ Rev.\ D {\bf 89}, 066017 (2014)
  [arXiv:1312.4542 [hep-th]].
  O.~Hohm and H.~Samtleben,
  ``Exceptional Form of D=11 Supergravity,''  Phys.\ Rev.\ Lett.\  {\bf 111}, 231601 (2013)  [arXiv:1308.1673 [hep-th]].  
  G.~Aldazabal, M.~Grana, D.~Marques and J.~A.~Rosabal,
  ``Extended geometry and gauged maximal supergravity,''  JHEP {\bf 1306}, 046 (2013)  [arXiv:1302.5419 [hep-th]].  
  G.~Aldazabal, M.~Graña, D.~Marqués and J.~é A.~Rosabal,
  ``The gauge structure of Exceptional Field Theories and the tensor hierarchy,''
  JHEP {\bf 1404}, 049 (2014)
  [arXiv:1312.4549 [hep-th]].
  O.~Hohm and H.~Samtleben,
  ``Exceptional Field Theory I: $E_{6(6)}$ covariant Form of M-Theory and Type IIB,''
  Phys.\ Rev.\ D {\bf 89}, 066016 (2014)
  [arXiv:1312.0614 [hep-th]].

\bibitem{Berman:2012vc}
  D.~S.~Berman, M.~Cederwall, A.~Kleinschmidt and D.~C.~Thompson,
  ``The gauge structure of generalised diffeomorphisms,''
  JHEP {\bf 1301}, 064 (2013)
  [arXiv:1208.5884 [hep-th]].

\bibitem{Hohm:2013jma}
  O.~Hohm and H.~Samtleben,
  ``U-duality covariant gravity,''  JHEP {\bf 1309}, 080 (2013)  [arXiv:1307.0509 [hep-th]].  
  O.~Hohm and H.~Samtleben,
  ``Gauge theory of Kaluza-Klein and winding modes,''  Phys.\ Rev.\ D {\bf 88}, 085005 (2013)  [arXiv:1307.0039 [hep-th]].  

\bibitem{Hohm:2013bwa}
  O.~Hohm, D.~Lust and B.~Zwiebach,
  ``The Spacetime of Double Field Theory: Review, Remarks, and Outlook,''  Fortsch.\ Phys.\  {\bf 61}, 926 (2013)  [arXiv:1309.2977 [hep-th]].  
  G.~Aldazabal, D.~Marques and C.~Nunez,
  ``Double Field Theory: A Pedagogical Review,''  Class.\ Quant.\ Grav.\  {\bf 30}, 163001 (2013)  [arXiv:1305.1907 [hep-th]].  
  D.~S.~Berman and D.~C.~Thompson,
  ``Duality Symmetric String and M-Theory,''  arXiv:1306.2643 [hep-th].  
  D.~S.~Berman, M.~Cederwall and M.~J.~Perry,
  ``Global aspects of double geometry,''
  arXiv:1401.1311 [hep-th].

\bibitem{Gualtieri:2003dx}
  M.~Gualtieri,
  ``Generalized complex geometry,''  math/0401221 [math-dg].  

\bibitem{Hitchin:2004ut}
  N.~Hitchin,
  ``Generalized Calabi-Yau manifolds,''  Quart.\ J.\ Math.\ Oxford Ser.\  {\bf 54}, 281 (2003)  [math/0209099 [math-dg]].  
  N.~Hitchin,
  ``Brackets, forms and invariant functionals,''  math/0508618 [math-dg].  
  M.~Gualtieri,
  ``Generalized complex geometry,''  math/0703298 [math.DG].  
  M.~Gualtieri,
  ``Generalized Kahler geometry,''
  arXiv:1007.3485 [math.DG].
  N.~Hitchin,
  ``Lectures on generalized geometry,''
  arXiv:1008.0973 [math.DG].
  G.~Cavalcanti,
  ``Introduction to generalized complex geometry,''
  Publicationes Matematicas do IMPA. (2007)
  P.~Koerber,
  ``Lectures on Generalized Complex Geometry for Physicists,''
  Fortsch.\ Phys.\  {\bf 59}, 169 (2011)
  [arXiv:1006.1536 [hep-th]].
  G.~R.~Cavalcanti and M.~Gualtieri,
  ``Generalized complex geometry and T-duality,''
  arXiv:1106.1747 [math.DG].
  H.~Bursztyn, G.~R.~Cavalcanti and M.~Gualtieri,
  ``Generalized Kaehler geometry of instanton moduli spaces,''
  arXiv:1203.2385 [math.DG].
  N.~Hitchin,
  ``Instantons, Poisson structures and generalized Kahler geometry,''
  Commun.\ Math.\ Phys.\  {\bf 265}, 131 (2006)
  [math/0503432 [math-dg]].
  N.~Hitchin,
  ``Generalized holomorphic bundles and the B-field action,''
  J.\ Geom.\ Phys.\  {\bf 61}, 352 (2011)
  [arXiv:1010.0207 [math.DG]].

\bibitem{Gualtieri:2007bq}
  M.~Gualtieri,
  ``Branes on Poisson varieties,''
  arXiv:0710.2719 [math.DG].
  M.~Boucetta,
  ``Riemannian geometry of Lie algebroids,''
  J.\ Egyptian Math.\ Soc., {\bf 19}, 2, 57-70. (2011)

\bibitem{Bressler:2005}
  P.~Bressler,
  ``The first Pontryagin class,''
  arXiv: math/0509563. (2005)
  Z.-J.~Liu, A.~Weinstein and P.~Xu,
  ``Manin triples for Lie bialgebroids,''
  J.\ Diff.\ Geom., {\bf 45}, 547-574. (1997)
  I.~Vaisman,
  ``Transitive Courant algebroids,''
  Int.\ J.\ Math.\ Math.\ Sci., {\bf 11}, 1737-1758. (2005)

\bibitem{Bursztyn:2005vwa}
  H.~Bursztyn, G.~R.~Cavalcanti and M.~Gualtieri,
  ``Reduction of Courant algebroids and generalized complex structures,''
  Adv.\ Math.\  {\bf 211}, 726 (2007)
  [math/0509640 [math.DG]].

\bibitem{Baraglia:2011dg}
  D.~Baraglia,
  ``Leibniz algebroids, twistings and exceptional generalized geometry,''
  J.\ Geom.\ Phys.\  {\bf 62}, 903 (2012)
  [arXiv:1101.0856 [math.DG]].
  R.~Rubio,
  ``Bn-generalized geometry and G2(2)-structures,''
  J.\ Geom.\ Phys.\  {\bf 73}, 150-156 (2013)
  [arXiv:1301.3330 [math.DG]].

\bibitem{Hull:2007zu}
  C.~M.~Hull,
  ``Generalised Geometry for M-Theory,''  JHEP {\bf 0707}, 079 (2007)  [hep-th/0701203].  
  P.~P.~Pacheco and D.~Waldram,
  ``M-theory, exceptional generalised geometry and superpotentials,''  JHEP {\bf 0809}, 123 (2008)  [arXiv:0804.1362 [hep-th]].  
  C.~Hillmann,
  ``Generalized E(7(7)) coset dynamics and D=11 supergravity,''  JHEP {\bf 0903}, 135 (2009)  [arXiv:0901.1581 [hep-th]].  
  D.~S.~Berman and M.~J.~Perry,
  ``Generalized Geometry and M theory,''  JHEP {\bf 1106}, 074 (2011)  [arXiv:1008.1763 [hep-th]].  

\bibitem{Wu:2013sha}
  H.~Wu and H.~Yang,
  ``Double Field Theory Inspired Cosmology,''
  arXiv:1307.0159 [hep-th].

\bibitem{Wu:2013ixa}
  H.~Wu and H.~Yang,
  ``New Cosmological Signatures from Double Field Theory,''
  arXiv:1312.5580 [hep-th].
  J.~Berkeley, D.~S.~Berman and F.~J.~Rudolph,
  ``Strings and Branes are Waves,''
  arXiv:1403.7198 [hep-th].

\bibitem{Kugo:1992md}
  T.~Kugo and B.~Zwiebach,
  ``Target space duality as a symmetry of string field theory,''
  Prog.\ Theor.\ Phys.\  {\bf 87}, 801 (1992)
  [hep-th/9201040].

\bibitem{Betz:2014aia}
  A.~Betz, R.~Blumenhagen, D.~Lust and F.~Rennecke,
  ``A Note on the CFT Origin of the Strong Constraint of DFT,''
  arXiv:1402.1686 [hep-th].

\bibitem{Ho:2013iia}
  P.~-M.~Ho,
  ``Gauge Symmetries from Nambu-Poisson Brackets,''
  Universe {\bf 1}, no. 4, 46 (2013).
  P.~-M.~Ho and C.~-T.~Ma,
  ``S-Duality for D3-Brane in NS-NS and R-R Backgrounds,''
  arXiv:1311.3393 [hep-th].
  P.~-M.~Ho and C.~-T.~Ma,
  ``Effective Action for Dp-Brane in Large RR (p-1)-Form Background,''
  JHEP {\bf 1305}, 056 (2013)
  [arXiv:1302.6919 [hep-th]].
  P.~-M.~Ho and Y.~Matsuo,
  ``Note on non-Abelian two-form gauge fields,''
  JHEP {\bf 1209}, 075 (2012)
  [arXiv:1206.5643 [hep-th]].
  P.~-M.~Ho, K.~-W.~Huang and Y.~Matsuo,
  ``A Non-Abelian Self-Dual Gauge Theory in 5+1 Dimensions,''
  JHEP {\bf 1107}, 021 (2011)
  [arXiv:1104.4040 [hep-th]].
  P.~-M.~Ho and C.~-H.~Yeh,
  ``D-brane in R-R Field Background,''
  JHEP {\bf 1103}, 143 (2011)
  [arXiv:1101.4054 [hep-th]].
  C.~-H.~Chen, K.~Furuuchi, P.~-M.~Ho and T.~Takimi,
  ``More on the Nambu-Poisson M5-brane Theory: Scaling limit, background independence and an all order solution to the Seiberg-Witten map,''
  JHEP {\bf 1010}, 100 (2010)
  [arXiv:1006.5291 [hep-th]].
  P.~-M.~Ho,
  ``A Concise Review on M5-brane in Large C-Field Background,''
  Chin.\ J.\ Phys.\  {\bf 48}, 1 (2010)
  [arXiv:0912.0445 [hep-th]].
  P.~-M.~Ho, Y.~Imamura, Y.~Matsuo and S.~Shiba,
  ``M5-brane in three-form flux and multiple M2-branes,''
  JHEP {\bf 0808}, 014 (2008)
  [arXiv:0805.2898 [hep-th]].
  P.~-M.~Ho and Y.~Matsuo,
  ``M5 from M2,''
  JHEP {\bf 0806}, 105 (2008)
  [arXiv:0804.3629 [hep-th]].
  C.~-T.~Ma and C.~-H.~Yeh,
  ``Supersymmetry and BPS States on D4-brane in Large C-field Background,''
  JHEP {\bf 1303}, 131 (2013)
  [arXiv:1210.4191 [hep-th]].
  P.~-M.~Ho, C.~-T.~Ma and C.~-H.~Yeh,
  ``BPS States on M5-brane in Large C-field Background,''
  JHEP {\bf 1208}, 076 (2012)
  [arXiv:1206.1467 [hep-th]].
  P.~Schupp and B.~Jurco,
  ``Nambu Sigma Model and Branes,''
  PoS CORFU {\bf 2011}, 045 (2011)
  [arXiv:1205.2595 [hep-th]].
  B.~Jurco and P.~Schupp,
  ``Nambu-Sigma model and effective membrane actions,''
  Phys.\ Lett.\ B {\bf 713}, 313 (2012)
  [arXiv:1203.2910 [hep-th]].
  B.~Jurco, P.~Schupp and J.~Vysoky,
  ``Nambu-Poisson Gauge Theory,''
  Physics Letters B 733C (2014), pp. 221-225
  [arXiv:1403.6121 [hep-th]].

\bibitem{Lawrence:2006ma}
  A.~Lawrence, M.~B.~Schulz and B.~Wecht,
  ``D-branes in nongeometric backgrounds,''
  JHEP {\bf 0607}, 038 (2006)
  [hep-th/0602025].
  C.~Albertsson, T.~Kimura and R.~A.~Reid-Edwards,
  ``D-branes and doubled geometry,''
  JHEP {\bf 0904}, 113 (2009)
  [arXiv:0806.1783 [hep-th]].
  C.~Albertsson, S.~-H.~Dai, P.~-W.~Kao and F.~-L.~Lin,
  ``Double Field Theory for Double D-branes,''
  JHEP {\bf 1109}, 025 (2011)
  [arXiv:1107.0876 [hep-th]].
  S.~Kawai and Y.~Sugawara,
  ``D-branes in T-fold conformal field theory,''
  JHEP {\bf 0802}, 027 (2008)
  [arXiv:0709.0257 [hep-th]].

\bibitem{Asakawa:2012px}
  T.~Asakawa, S.~Sasa and S.~Watamura,
  ``D-branes in Generalized Geometry and Dirac-Born-Infeld Action,''
  JHEP {\bf 1210} (2012) 064
  [arXiv:1206.6964 [hep-th]].
  T.~Asakawa, H.~Muraki and S.~Watamura,
  ``D-brane on Poisson manifold and Generalized Geometry,''
  arXiv:1402.0942 [hep-th].
  B.~Jurco, P.~Schupp and J.~Vysoky,
  ``On the Generalized Geometry Origin of Noncommutative Gauge Theory,''
  JHEP {\bf 1307}, 126 (2013)
  [arXiv:1303.6096 [hep-th]].
  B.~Jurco, P.~Schupp and J.~Vysoky,
  ``Extended generalized geometry and a DBI-type effective action for branes ending on branes,''
  arXiv:1404.2795 [hep-th].

\bibitem{Berman:2011kg}
  D.~S.~Berman, E.~T.~Musaev and M.~J.~Perry,
  ``Boundary Terms in Generalized Geometry and doubled field theory,''
  Phys.\ Lett.\ B {\bf 706}, 228 (2011)
  [arXiv:1110.3097 [hep-th]].
\end{thebibliography}
\end{document}